\documentclass[preprint,12pt]{elsarticle}

\usepackage{amssymb}
\usepackage{amsthm}

\journal{Journal of Computational Physics}

\usepackage{graphicx}
\usepackage[utf8]{inputenc}
\usepackage[colorlinks=true]{hyperref}
\usepackage{amsmath}
\usepackage{subfig}
\usepackage[left=2.5cm,right=2.5cm]{geometry}
\usepackage[english]{babel}

\usepackage[toc,page]{appendix} 
\usepackage{tabularx} 

\usepackage{tikz}
\usetikzlibrary{shapes,arrows}
\tikzset{font=\small}

\usepackage{pgfplots}
\pgfplotsset{compat=newest}
\usepgfplotslibrary{groupplots}

\definecolor{myblue}{rgb}{0.050980,0.34118,0.69804}
\definecolor{myred}{rgb}{1,0.21961,0.28627}
\definecolor{mygrey}{rgb}{0.43137,0.43137,0.43137}
\definecolor{mygreen}{rgb}{0.1211,0.7304,0.2304}
\definecolor{myblue_changes}{RGB}{18,87,166}

\usepackage{enumerate}
\usepackage[mathscr]{euscript}
\usepackage[T1]{fontenc}
\usepackage{caption}
\usepackage{multirow}

\theoremstyle{plain}

\begin{document}
\begin{frontmatter}





\title{Multi-stage splitting integrators for sampling with modified Hamiltonian Monte Carlo methods}



\author[bcam]{Tijana Radivojevi\'c\corref{cor}}
\author[bcam]{Mario Fern\'andez-Pend\'as}
\author[carlosiii]{Jes\'us Mar\'ia Sanz-Serna}
\author[bcam,ikerbasque]{Elena Akhmatskaya}

\cortext[cor]{Contact address: \texttt{tijana.radivojevic@gmail.com}}

\address[bcam]{BCAM - Basque Center for Applied Mathematics, Alameda de Mazarredo 14, E-48009 Bilbao, Spain}
\address[ikerbasque]{IKERBASQUE, Basque Foundation for Science, Mar\'ia D\'iaz de Haro 3, E-48013 Bilbao, Spain}
\address[carlosiii]{Departamento de Matemáticas, Universidad Carlos III de Madrid, Avenida de la Universidad 30,\\ E-28911 Legan\'es (Madrid), Spain}

\begin{abstract}
Modified Hamiltonian Monte Carlo (MHMC) methods combine the ideas behind two popular sampling approaches: Hamiltonian Monte Carlo (HMC) and importance sampling. As in the HMC case, the bulk of the computational cost of MHMC algorithms lies in the numerical integration of a Hamiltonian system of differential equations. We suggest novel integrators designed to \textcolor{black}{enhance accuracy and sampling performance of MHMC methods.} The novel integrators belong to  families of splitting algorithms and are therefore easily implemented. We identify optimal integrators within the families by minimizing  the energy error or the average energy error. We derive and discuss in detail the modified Hamiltonians of the new integrators, as the evaluation of those Hamiltonians is key to the efficiency of the overall algorithms. Numerical experiments show that the use of the new integrators may improve very significantly the sampling performance of MHMC methods,  in both statistical and molecular dynamics problems.
 \end{abstract}


\end{frontmatter}






\section{Introduction}

Hybrid/Hamiltonian Monte Carlo (HMC) \cite{DKPR87,Neal10} and importance sampling \cite{KahnMarshall53}
 algorithms have been effectively used for sampling in molecular simulation and computational statistics.
 Successful alternatives to these two approaches may be found within the class of modified Hamiltonian Monte Carlo (MHMC) methods that combine HMC and importance sampling, as suggested in \cite{IH04,Akhmatskaya08,Radivojevic:2017}.
  Taking advantage of the fact that symplectic integrators preserve modified
   Hamiltonians more accurately than they conserve the true Hamiltonian, the authors of those references
    proposed to sample with respect to modified/shadow Hamiltonians and to recover the target distribution by reweighting.
  The resulting algorithms are capable of maintaining high acceptance rates and usually exhibit better efficiency than their predecessor HMC, that samples with respect to the true Hamiltonian \cite{Radivojevic16,Akhmatskaya:2012,Wee08,Escribano17,Radivojevic:2017}.

  The first methods of the MHMC class were derived for atomistic simulations and differed from each other in the way of performing the three main components of the algorithm---momentum updates, computation of modified Hamiltonians and integration of the Hamiltonian dynamics.
For example, in the (Separable) Shadow Hybrid Monte Carlo methods introduced in \cite{IH04,SHSI09} a full  momentum update is used, whereas in  Targeted Shadow Hybrid Monte Carlo \cite{AR06} and  Generalized Shadow Hybrid Monte Carlo (GSHMC \cite{Akhmatskaya08}) suitable modifications of the partial momentum update of \cite{Horowitz91} are advocated in order to better mimic the dynamics and to enhance sampling.
More recent MHMC methods  aim at specific applications such as computational statistics (Mix\&Match Hamiltonian Monte Carlo \cite{Radivojevic:2017}), or, in molecular problems, multi-scale (MTS-GSHMC \cite{Escribano14}), meso-scale (meso-GSHMC \cite{Akhmatskaya:2011}) and solid-state (RSM-GSHMC \cite{Escribano17}) simulations.
As demonstrated in the original papers, for particular problems, the use of MHMC methods resulted in a sampling efficiency several times higher than that observed with  conventional sampling techniques, such as Molecular Dynamics, Monte Carlo (MC) and HMC.

Two elements critically affect the efficiency of  MHMC methods. On the one hand, the calculation of modified Hamiltonians at every Monte Carlo step introduces  computational overheads when compared to HMC, and hence finding efficient ways of calculating modified Hamiltonians is of interest. On the other hand, the accuracy of the numerical integrator directly influences the acceptance rate of the generated proposals and therefore the sampling efficiency. Verlet/leapfrog has been the integrator of choice for  MHMC methods and until recently such a choice has never been challenged. On the contrary, the search for the best performing integrator for \textcolor{black}{Hamiltonian problems and especially for} HMC has been a live research topic for a decade \cite{McL95, BCSS14,CSMS15,FPASS16,Bou-Rabee:2018}. In particular, it has been demonstrated that replacing in HMC the standard Verlet integrator  with a splitting integrator that includes a specifically chosen parameter value may significantly improve, for a range of step sizes, the conservation of the Hamiltonian and thus the \textcolor{black}{overall performance of the algorithm} \cite{McL95,BCSS14}.

The goal of this paper is to adapt to MHMC methods the splitting schemes introduced in the literature
as successful alternatives to Verlet in the HMC case. In particular, we propose  novel multi-stage integrators
 that may improve the conservation of modified Hamiltonians \textcolor{black}{for a range of time steps} and, as a
 consequence \textcolor{black}{the sampling performance
  in the MHMC methods.}
These integrators are characterized by  parameter values resulting from minimization of the (expected) Hamiltonian error introduced by integration.
More specifically, we derive  two- and three-stage integrators along with their  corresponding modified Hamiltonians, and investigate their performance within two specific MHMC methodologies, GSHMC \cite{Akhmatskaya08} and MMHMC \cite{Radivojevic:2017}, developed for molecular simulation and computational statistics, respectively.

The paper is structured as follows.
In Section~\ref{sec:MHMC} we present a brief review of  modified HMC methods and their main components.
In Section~\ref{Sec:Mod_Ham} we provide new formulations of modified Hamiltonians of 4th and 6th order for splitting integrating schemes that include families of two- and three-stage integrators.
Section~\ref{Sec:Integrators} provides  new, optimized, multi-stage integrators arising from two different optimization approaches, namely (i) minimization, in the limit where the step-size approaches 0, of the error in modified Hamiltonian introduced by numerical integration, and (ii) minimization, for finite step-sizes, of the expected value of that error.
Section~\ref{Sec:Experiments} is devoted to testing the new integrators. We give details of the methods and models tested and the  performance metrics employed, together with results of comparisons between the approaches suggested here and other popular integration techniques for HMC methods.
Our conclusions are summarized in Section~\ref{Sec:Conclusions}.

\section{Modified Hamiltonian Monte Carlo methods}\label{sec:MHMC}

The goal of a modified Hamiltonian Monte Carlo (MHMC) method is to estimate an integral
\begin{equation}\label{eq:integral}
I=\int f(\mathbf x) \pi(\mathbf x)\mathrm d \mathbf x,
\end{equation}
i.e.\ the expected value of a function of interest $f$ with respect to the density $\pi(\mathbf x)$, known up to a multiplicative constant, of a position variable $\mathbf x$. Instead of sampling from the canonical distribution
\begin{equation}\label{eq:Distribution}
\pi(\mathbf x,\mathbf p)\propto\exp{\left(-\beta H(\mathbf x,\mathbf p)\right)}
\end{equation}
($\beta$ is the inverse temperature, $H$ the Hamiltonian and  $\mathbf p$ the auxiliary momentum variable), as is the case of general HMC algorithms,  MHMC methods sample from an importance canonical density
\begin{equation}\label{eq:hk}
\tilde{\pi}(\mathbf x,\mathbf p)\propto \exp{\left(-\beta \tilde{H}^{[k]}(\mathbf x,\mathbf p)\right)},
\end{equation}
where  $\tilde{H}^{[k]}$ denotes the $k$th order truncation of the modified Hamiltonian \(\tilde H\) that is preserved exactly by the symplectic integrator under consideration.

In this paper we consider  separable Hamiltonians
\begin{equation}\label{eq:Ham}
H(\mathbf x,\mathbf p)=U(\mathbf x)+\frac 1 2 \mathbf p^T M^{-1}\mathbf p,
\end{equation}
with $M$  a symmetric positive definite matrix (mass matrix ) and $U$ the potential function. The dynamics associated to this Hamiltonian is governed by the system of ordinary differential equations
\begin{equation}\label{eq:HD}
\frac{\mathrm d \mathbf x}{\mathrm d t}=M^{-1} \mathbf p, \; \; \frac{\mathrm d \mathbf p}{\mathrm d t}=-U_{\mathbf x}(\mathbf x).
\end{equation}

We describe now a generic algorithm for an MHMC method. Given a sample $(\mathbf x,\mathbf p)$ from the distribution ${\tilde{\pi}}$, the next sample $(\mathbf x^{\mbox{\small new}},\mathbf p^{\mbox{\small new}})$ is determined as follows:
\begin{itemize}
	\item Define the new momentum $\mathbf p^*$ by applying a momentum refreshment procedure that preserves the importance density  ${\tilde{\pi}}$ in \eqref{eq:hk}.
	\item Generate a proposal  $(\mathbf x',\mathbf p')$ by simulating, using a symplectic and reversible numerical integrator, the Hamiltonian system \eqref{eq:HD} with initial condition $(\mathbf x,\mathbf p^*)$.
	\item Accept the proposal as the next sample $(\mathbf x^{\mbox{\small new}},\mathbf p^{\mbox{\small new}})$ with  probability
	\begin{equation*}
		\quad \alpha=\min \left\{1,\frac{\tilde{\pi}(\mathbf x',\mathbf p')}{\tilde{\pi}(\mathbf x,\mathbf p^*)}\right\}.
	\end{equation*}
	Otherwise, set $(\mathbf x^{\mbox{\small new}},\mathbf p^{\mbox{\small new}})$ to $(\mathbf x, -\mathbf p^*)$, i.e., flip the momentum.
\end{itemize}

 Once the samples have been obtained using MHMC, importance reweighting is required in order to estimate \eqref{eq:integral}, since the samples are generated with respect to the importance density \eqref{eq:hk} on the joint state space. The desired distribution $\pi(\mathbf x)$ is recovered by marginalizing momenta variables from \eqref{eq:Distribution}.
If $f_n=f(\mathbf x^n)$, $n = 1, 2, \ldots, N$, are the values of an observable along a sequence of states $(\mathbf x^n, \mathbf p^n)$ drawn from $\tilde{\pi}$, \eqref{eq:integral} is estimated as
\begin{equation}\label{eq:Reweighting}
	\hat I = \frac{\sum_{n=1}^N w_n f_n}{\sum_{n=1}^N w_n},
\end{equation}
where the \textit{importance weights} are given by
\begin{equation*}
	w_n =\exp{\left(-\beta\big(H(\mathbf x^n, \mathbf p^n)-\tilde{H}^{[k]}(\mathbf x^n, \mathbf p^n)\big)\right)}.
\end{equation*}

\section{Modified Hamiltonians for splitting integrators}\label{Sec:Mod_Ham}

Our intention is to use within  MHMC methods  numerical integrators that may offer better conservation properties than the commonly used Verlet/leapfrog integrator. More specifically, we are interested in numerical integrators belonging to the two-stage
\begin{equation}\label{2S}
	\psi_h=\varphi^B_{bh}\circ\varphi^A_{\frac{h}{2}}\circ \varphi^B_{(1-2b)h}\circ\varphi^A_{\frac{h}{2}}\circ \varphi^B_{bh}
\end{equation}
and three-stage
\begin{equation}\label{3S}
	\psi_h=\varphi^B_{bh}\circ\varphi^A_{a h}\circ \varphi^B_{(\frac{1}{2}-b)h}\circ\varphi^A_{(1-2a)h}\circ \varphi^B_{(\frac{1}{2}-b)h}\circ\varphi^A_{a h}\circ \varphi^B_{bh}
\end{equation}
families of splitting methods, that require two or three gradient evaluations per step, respectively. The exact flows $\varphi^A_{h}$ and $\varphi^B_{h}$ are solutions to the split systems
\begin{equation}\label{HamA}
	A: \frac{\mathrm d \mathbf x}{\mathrm d t} = 0, \; \;
	\frac{\mathrm d \mathbf p}{\mathrm d t} = -U_{\mathbf x}(\mathbf x),
\end{equation}
and
\begin{equation}\label{HamB}
	B: \frac{\mathrm d \mathbf x}{\mathrm d t} = M^{-1} \mathbf p, \; \;
	\frac{\mathrm d \mathbf p}{\mathrm d t} = 0,
\end{equation}
respectively, corresponding to the Hamiltonian \eqref{eq:Ham}. A specific integrator  is fully characterized by the choice  of values for the coefficients $\{b\}$ or $\{a,b\}$.
\textcolor{black}{While it is possible to swap the roles of the kinetic and potential energies in the splitting algorithms, in this study we only consider integrators starting with an update of momenta for reasons discussed in detail in \cite{Bou-Rabee:2018}.}

We point out that for the specific choice $b=1/4$, the application of \(\psi_h\) in \eqref{2S} yields the  same result as performing two consecutive Verlet steps each with step size $h/2$. Similarly for $b= 1/6$,  $a=1/3$, \eqref{3S} is equivalent to three Verlet steps of size $h/3$. It is also important to emphasize that since one step of a splitting integrator consists of a sequence of flows of \eqref{HamA}--\eqref{HamB} (some times called kicks and drifts respectively), the implementation of methods of the families \eqref{2S}--\eqref{3S} is very similar to that of the Verlet algorithm. Therefore it is easy to incorporate splitting algorithms to existing software based on the Verlet integrator.

To use splitting integrators within an MHMC method,  appropriate modified Hamiltonians are required.
For the Verlet integrator, one procedure to compute modified Hamiltonians of orders up to 24 is provided in \citep{Skeel01,Engle05}.
It is further improved in \citep{MoanNiesen10} using Richardson extrapolation. That approach could be generalized   to  multi-stage integrators. However, it requires a modification of the integrator by introducing into the dynamics an additional scalar variable.
We therefore opt here for a different strategy. For the families of splitting integrators above, we begin by writing the expansion of the modified Hamiltonian  in terms of Poisson brackets of the partial Hamiltonians $A$ and $B$ of the split systems \eqref{HamA}-\eqref{HamB}. We recall that the \emph{Poisson bracket} of two functions $F,G: \mathbb R^{2D}\rightarrow \mathbb R$ is defined as
	\begin{equation*}\label{PoissBr_def}
		\{F,G \}(\mathbf z)=F_{\mathbf z}(\mathbf z)^T \mathbf J G_{\mathbf z}(\mathbf z), \;  \mathbf J = \begin{bmatrix}
			0 & I \\
			-I & 0 \\
		\end{bmatrix}.
	\end{equation*}
The expression for the modified Hamiltonian $\tilde H$ is found to be
\begin{eqnarray}\label{modifiedHam}
	\tilde{H} = H &+&h^2 \alpha \{A,A,B \} + h^2 \beta \{B,B,A \}  \nonumber \\
	&+& h^4 \gamma_{1}\{A,A,A,A,B \} + h^4 \gamma_{2} \{ B,A,A,A,B\}  \\
	&+& h^4 \gamma_{3} \{B,B,A,A,B \} +  h^4 \gamma_{4} \{A,A,B,B,A \} +\mathcal O(h^6)\nonumber,
\end{eqnarray}
where $\alpha,\beta,\gamma_{1-4}$ are polynomials in  the  integrator  coefficients $b$ or $a,b$ \citep{BCSS14}
and expressions such as $\{A,A,B\}$ refer to iterated Poisson brackets $\{A,\{A,B\}\}$.

The expressions for the expansion of $\tilde{H}$ to arbitrarily high order may be obtained by directly
applying the Baker-Campbell-Hausdorff (BCH) formula to the exponentials of the Lie derivatives
of the partial Hamiltonians $A$ and $B$. Recall that the \emph{Lie derivative} with respect to a function $F:
\mathbb R^{2D}\rightarrow \mathbb R$ is defined in terms of Poisson bracket as
\begin{equation}\label{LieDer_def}
	\mathcal L_F(\cdot)=\{\cdot,F\}.
\end{equation}
\textcolor{black}{Unfortunately computations based  on the BCH are cumbersome if the order is high \citep{SanzSerna94}.
Alternatively, for symmetric composition methods, the coefficients multiplying the Poisson brackets for the
4th, 6th and 8th order truncations of the modified Hamiltonian  can be derived from expressions presented in
\cite{OMF02}.}

Here we propose two alternative ways to derive the expression for the 4th and 6th order modified Hamiltonians. The first uses derivatives of the potential function, obtained either by automatic differentiation or from analytical expressions, whereas the second relies on numerical time derivatives of the gradient, obtained through  quantities available along the simulation.

\subsection{Modified Hamiltonians in terms of derivatives of the potential}
For problems in which derivatives of the potential function can be computed, either from analytical
expressions or using automatic differentiation, we derive the 4th and 6th order modified Hamiltonians by first
expanding the iterated brackets in \eqref{modifiedHam} using  the definition of the Poisson bracket. This
yields
\begin{eqnarray*}
	\{A,A,B \} &=& \mathbf p^T M^{-1}U_{\mathbf x\mathbf x}(\mathbf x) M^{-1}\mathbf p\\
	\{B,B,A \} &=& {U_{\mathbf x}(\mathbf x)}^T M^{-1}U_{\mathbf x}(\mathbf x)\\
	\{A,A,A,A,B \}&=&  U_{\mathbf x \mathbf x \mathbf x \mathbf x }(\mathbf x ) M^{-1}\mathbf pM^{-1}\mathbf p M^{-1}\mathbf p M^{-1}\mathbf p\\
	\{ B,A,A,A,B\}&=& -3 {U_{\mathbf x }(\mathbf x )}^T M^{-1}U_{\mathbf x \mathbf x \mathbf x}(\mathbf x) M^{-1}\mathbf pM^{-1}\mathbf p \nonumber\\
	\{B,B,A,A,B \}&=& 2{U_{\mathbf x }(\mathbf x)}^T M^{-1}U_{\mathbf x\mathbf x}(\mathbf x) M^{-1}U_{\mathbf x}(\mathbf x)\\
	\{A,A,B,B,A \}&=&  2 {U_{\mathbf x }(\mathbf x )}^T M^{-1}U_{\mathbf x \mathbf x \mathbf x}(\mathbf x) M^{-1}\mathbf pM^{-1}\mathbf p \\
	&&+ 2 \mathbf p^T M^{-1}U_{\mathbf x\mathbf x}(\mathbf x) M^{-1}U_{\mathbf x\mathbf x}(\mathbf x) M^{-1}\mathbf p
\end{eqnarray*}
and leads to the following  4th and 6th order modified Hamiltonians for splitting integrators
\begin{align}
	\tilde{H}^{[4]}(\mathbf x,\mathbf p)  = & H(\mathbf x,\mathbf p) +h^2 c_{21} \mathbf p^T M^{-1}U_{\mathbf x\mathbf x}(\mathbf x) M^{-1}\mathbf p+ h^2 c_{22} {U_{\mathbf x}(\mathbf x)}^T M^{-1}U_{\mathbf x}(\mathbf x),\label{modHam4_an}   \\
	\nonumber \\
	\tilde{H}^{[6]}(\mathbf x,\mathbf p) = & \tilde{H}^{[4]}(\mathbf x,\mathbf p) + h^4 c_{41} U_{\mathbf x \mathbf x \mathbf x \mathbf x }(\mathbf x ) M^{-1}\mathbf pM^{-1}\mathbf p M^{-1}\mathbf p M^{-1}\mathbf p  \label{modHam6_an} \\
	&+ h^4 c_{42}{U_{\mathbf x }(\mathbf x )}^T M^{-1}U_{\mathbf x \mathbf x \mathbf x}(\mathbf x) M^{-1}\mathbf pM^{-1}\mathbf p \nonumber\\
	&+ h^4 c_{43}{U_{\mathbf x }(\mathbf x)}^T M^{-1}U_{\mathbf x\mathbf x}(\mathbf x) M^{-1}U_{\mathbf x}(\mathbf x) \nonumber \\
	& +  h^4 c_{44} \mathbf p^T M^{-1}U_{\mathbf x\mathbf x}(\mathbf x) M^{-1}U_{\mathbf x\mathbf x}(\mathbf x) M^{-1}\mathbf p, \nonumber %
\end{align}
where
\begin{equation}\label{coeff_c_gamma}
	c_{21}=\alpha, \hspace{0.5cm} c_{22}=\beta,  \hspace{0.5cm} c_{41}=\gamma_1, \hspace{0.5cm} c_{42}=2\gamma_4-3\gamma_2, \hspace{0.5cm} c_{43}= 2 \gamma_3, \hspace{0.5cm} c_{44}=2\gamma_4.
\end{equation}
The coefficients $\alpha,\beta,\gamma_{1-4}$ can be derived from expressions  given in  \citep{OMF02} where the authors analyzed   so-called force-gradient integrators for molecular dynamics. In particular, they considered  splitting integrators that are extended by an additional higher-order operator into the single-exponen\-tial propagators.
If the potential function is quadratic, i.e.\ corresponding to problems of sampling from Gaussian distributions/harmonic oscillators, \textcolor{black}{$\{A,A,A,A,B \} = 0$ and $\{ B,A,A,A,B\} = 0$, and then} the 6th order modified Hamiltonian \eqref{modHam6_an} simplifies to
\begin{align}
	\tilde{H}^{[6]}(\mathbf x,\mathbf p) =  \tilde{H}^{[4]}(\mathbf x,\mathbf p)
	&+ h^4 c_{43}{U_{\mathbf x }(\mathbf x)}^T M^{-1}U_{\mathbf x\mathbf x}(\mathbf x) M^{-1}U_{\mathbf x}(\mathbf x) \nonumber  \\
	& +  h^4 c_{44} \mathbf p^T M^{-1}U_{\mathbf x\mathbf x}(\mathbf x) M^{-1}U_{\mathbf x\mathbf x}(\mathbf x) M^{-1}\mathbf p. \nonumber %
\end{align}

Combining \eqref{coeff_c_gamma} with the expressions for $\alpha,\beta,\gamma_{1-4}$, we obtain the following coefficients for the two-stage  integrator family \eqref{2S}
{\small
	\begin{equation}\label{modHam2S}
		\begin{aligned}
			c_{21}=& \frac{1}{24}\Big(6b-1\Big) \\
			c_{22}=& \frac{1}{12}\Big(6b^2-6b+1\Big)\\
			c_{41} =& \frac{1}{5760}\Big(7-30b\Big) \\
			c_{42} =& \frac{1}{240}\Big(-10b^2+15b-3\Big) \\
			c_{43} =& \frac{1}{120}\Big(-30b^3+35b^2-15b+2\Big) \\
			c_{44} =& \frac{1}{240}(20b^2-1).
		\end{aligned}
	\end{equation}
}
For three-stage integrators \eqref{3S} (a two-parameter family), we get
{\small
	\begin{equation}\label{modHam3S}
		\begin{aligned}
			c_{21}=& \frac{1}{12} \Big(1-6a(1-a)(1-2b)\Big)\\
			c_{22}=& \frac{1}{24} \Big(6a(1-2b)^2-1\Big)\\
			c_{41} =& \frac{1}{720}\Big( 1 + 2 (a-1) a (8 + 31 (a-1) a) (1 - 2 b) - 4 b\Big)\\
			c_{42}=& \frac{1}{240} \Big(6 a^3 (1 - 2 b)^2 -
			a^2 (19 - 116 b + 36 b^2 + 240 b^3)+ a (27 - 208 b + 308 b^2)- 48 b^2+ 48 b    -7\Big) \\
			c_{43} =&\frac{1}{180}\Big(1 + 15 a (1- 2 b) (-1 + 2 a (2 - 3 b + a (4 b-2)))\Big)  \\
			c_{44} =& \frac{1}{240} \Big(-1 + 20 a (1 - 2 b) (b + a (1 + 6 (b-1) b))\Big).
		\end{aligned}
	\end{equation}
}

Using \eqref{modHam2S} one can also obtain the modified Hamiltonian for the Verlet integrator, since, as
pointed out above,
 two steps of Verlet integration with step size $h/2$ are equivalent to one step with step size $h$ of the two-stage integrator with $b=1/4$. In this way the Verlet coefficients are found to be
\begin{eqnarray}\label{modHamVerlet}
	c_{21}&=& \frac{1}{12}, \hspace{0.5cm} c_{22}=- \frac{1}{24} \\
	c_{41}& = &-\frac{1}{720}, \hspace{0.5cm} c_{42} = \frac{1}{120}, \hspace{0.5cm}c_{43} = -\frac{1}{240}, \hspace{0.5cm} c_{44} =\frac{1}{60}. \nonumber
\end{eqnarray}

\begin{figure}[!ht]
	\centering
	\begin{tikzpicture}
		\begin{groupplot}[group style={group size=2 by 1, horizontal sep=1.7cm, vertical sep=2cm}]
			\nextgroupplot[width=8.1cm, height=5.67cm,
				grid = major,
				xmin=0, xmax=155, xlabel={number of integration steps}, xtick={0,50,100,150}, xticklabels={0,50,100,150},
				ymin=0, ymax=10, ylabel={overhead (\%)}, ytick={0,2,4,6,8,10}, yticklabels={0,2,4,6,8,10},
				legend pos=north east, legend style={draw=black,fill=white,legend cell align=left}]
	
				\addplot[color=mygrey,dashed,line width=2pt] table[x=L,y=five]{overheads.dat};
				\addlegendentry{$D = 500$};
	
				\addplot[color=mygrey,line width=2pt] table[x=L,y=two]{overheads.dat};
				\addlegendentry{$D = 2000$};

			\nextgroupplot[width=8.1cm, height=5.67cm,
				grid = major,
				xmin=-5, xmax=310, xlabel={number of integration steps}, xtick={0,100,200,300}, xticklabels={0,100,200,300},
				ymin=0, ymax=350, ylabel={overhead (\%)}, ytick={0,50,100,150,200,250,300,350}, yticklabels={0,50,100,150,200,250,300,350},
				legend pos=north east, legend style={draw=black,fill=white,legend cell align=left}]
	
				\addplot[color=mygrey,line width=2pt] table[x=L,y=sixty]{overheads2.dat};
				\addlegendentry{$D = 61$};
		\end{groupplot}
	\end{tikzpicture}
	\caption{Computational overhead of MMHMC compared to HMC for models with a tridiagonal (left) and a dense Hessian matrix (right) using the 4th order modified Hamiltonian \eqref{modHam4_an} with all required derivatives calculated analytically.}
	\label{Fig:overheads_modHam4A}
\end{figure}
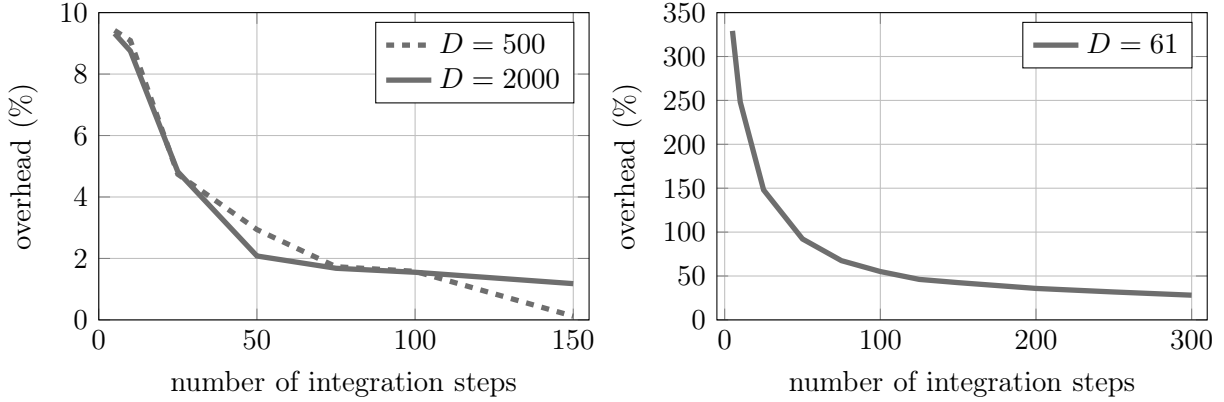

Figure \ref{Fig:overheads_modHam4A} illustrates, for two-stage integrators, the computational overheads, with respect to the HMC method, of MMHMC based on the 4th order modified Hamiltonian \eqref{modHam4_an}. The left-hand graph presents the overhead for a model with a tridiagonal Hessian matrix $U_{\mathbf x\mathbf x}(\mathbf x)$, as appeared in the stochastic volatility model \cite{Radivojevic16}, and indicates that, for two different dimensions $D$ of the system, the overhead becomes negligible as the number of integration steps increases. In contrast, for models with a dense Hessian matrix, as appeared in the Bayesian logistic regression model \cite{Radivojevic16}, computation of modified Hamiltonians may introduce a significant additional cost, as shown in the right-hand graph.

\subsection{Modified Hamiltonians in terms of time derivatives of the gradient}\label{Sec:modHam_num}

For applications where the potential function has a dense Hessian matrix (and tensors of higher derivatives),
the
 overhead resulting from the computation of modified Hamiltonians reduces the advantages of MHMC methods.
In order to implement such computations in an efficient manner, we propose to express modified Hamiltonians in terms of quantities that are available during the simulation, as  done in \cite{Akhmatskaya08}.  However, instead of making use of  time derivatives of the position vectors, as  in the original GSHMC method  \cite{Akhmatskaya08}, here  we employ time derivatives of the gradient of the potential function, as follows (superscripts in brackets indicate the order of the time derivative)
\begin{eqnarray}
	\label{U_der_identities}
	U_{\mathbf x}^{(1)} &=& U_{\mathbf x \mathbf x }(\mathbf x ) M^{-1}\mathbf p \nonumber \\
	U_{\mathbf x }^{(2)} &=& U_{\mathbf x \mathbf x \mathbf x }(\mathbf x ) M^{-1}\mathbf pM^{-1}\mathbf p-U_{\mathbf x \mathbf x }(\mathbf x ) M^{-1}U_{\mathbf x }(\mathbf x )\\
	U_{\mathbf x }^{(3)} &=&U_{\mathbf x \mathbf x \mathbf x \mathbf x }(\mathbf x ) M^{-1}\mathbf pM^{-1}\mathbf p M^{-1}\mathbf p -3 U_{\mathbf x \mathbf x \mathbf x }(\mathbf x ) M^{-1} U_{\mathbf x }(\mathbf x )M^{-1}\mathbf p \nonumber\\
	&&  - U_{\mathbf x \mathbf x }(\mathbf x ) M^{-1}U_{\mathbf x \mathbf x}(\mathbf x ) M^{-1}\mathbf p.\nonumber
\end{eqnarray}
Substituting the time derivatives \eqref{U_der_identities} into the analytical expressions
\eqref{modHam4_an}-\eqref{modHam6_an} for the 4th and 6th order modified Hamiltonians, respectively, one
obtains
\begin{eqnarray}
	\tilde{H}^{[4]}(\mathbf x ,\mathbf p)  &=& H(\mathbf x ,\mathbf p) +h^2 k_{21} \mathbf p^T M^{-1}{U_{\mathbf x }}^{(1)} + h^2 k_{22} {U_{\mathbf x }(\mathbf x)}^T M^{-1}U_{\mathbf x }(\mathbf x ), \label{sH4}  \\
	\nonumber \\
	\tilde{H}^{[6]}(\mathbf x ,\mathbf p) &=& \tilde{H}^{[4]}(\mathbf x ,\mathbf p) + h^4 k_{41}\mathbf p^T M^{-1}{U_{\mathbf x }}^{(3)} + h^4 k_{42} {{U_{\mathbf x }}(\mathbf x )}^TM^{-1}{U_{\mathbf x }}^{(2)} \label{sH6} \\
	&+& h^4 k_{43} {{U_{\mathbf x }}^{(1)}}^TM^{-1}{U_{\mathbf x}}^{(1)} +  h^4 k_{44} U_{\mathbf x }(\mathbf x )^TM^{-1} U_{\mathbf x \mathbf x }(\mathbf x ) M^{-1}U_{\mathbf x }(\mathbf x), \nonumber
\end{eqnarray}
where the coefficients are
\begin{eqnarray}\label{k_coeff}
	k_{21}&=&c_{21}, \hspace{0.5cm} k_{22}=c_{22}, \\
	k_{41}&=&c_{41}, \hspace{0.5cm} k_{42}=3c_{41} +c_{42}, \hspace{0.5cm} k_{43}=c_{41} +c_{44}, \hspace{0.5cm}  k_{44}=3c_{41}+c_{42}+c_{43}.\nonumber
\end{eqnarray}
We note that the  expression \eqref{sH4} does not include the Hessian of the potential and thus, allows  the computation of $\tilde{H}^{[4]}$ using quantities available from the integration of the dynamics.
However, this is not the case for the  6th order Hamiltonians. The last term in \eqref{sH6}, arising from an expansion of the Poisson bracket $\{B,B,A,A,B \}$, cannot be computed using time derivatives of available quantities and requires explicit calculation of the Hessian matrix of the potential function. Only for the Verlet integrator does this term vanish and the resulting coefficients are
\begin{eqnarray*}
	k_{21}&=& \frac{1}{12}, \hspace{0.5cm} k_{22}=- \frac{1}{24},\nonumber \\
	k_{41}& = &-\frac{1}{720}, \hspace{0.5cm} k_{42} = \frac{1}{240}, \hspace{0.5cm} k_{43} = \frac{11}{720}, \hspace{0.5cm} k_{44} =0.
\end{eqnarray*}
One can now write explicit expressions for the coefficients $k_{ij}$ by  substituting the expressions for the coefficients $c_{ij}$ \eqref{modHam2S} or   \eqref{modHam3S}  into  \eqref{k_coeff} for two- and three-stage integrators, respectively.

In the original GSHMC method, an interpolating polynomial of the positions $\mathbf x(t_i)= \mathbf x^i, \; i=n-k,\dots,n,\dots,n+k, \; n\in \{0,L\}$ is constructed from a numerical trajectory $\{\mathbf x^i\}^{L+k}_{i=-k}$, where $k=2$ and $k=3$ for the 4th and 6th order modified Hamiltonian, respectively. This requires four or six additional gradient calculations in order to compute $\tilde{H}^{[4]}$ or $\tilde{H}^{[6]}$, respectively.
Here we choose a different strategy and calculate the polynomial in terms of the gradient of the potential function
$$\mathbf U(t_i)= U_{\mathbf x}(\mathbf x^i), \; i=n-k,\dots,n,\dots,n+k.$$
With this approach,  $k=1$ for the 4th order and $k=2$ for the 6th order modified Hamiltonian, which implies that the evaluation of $\tilde{H}^{[4]}$ and $\tilde{H}^{[6]}$ requires fewer additional gradient computations, namely two and four, respectively. Note that $k$  corresponds to a multiple of the full integration step only in the case of the Verlet integrator; for other integrators it is the number of stages performed (e.g.\ $k=2$ corresponds to a half integration step of a four-stage method).
Also note that an efficient implementation does not include the unnecessary integration sub-step that would update momentum at the very beginning and very end of the numerical trajectory $\{U_{\mathbf x}(\mathbf x^i)\}_{i=-k}^{L+k}$.

Time derivatives of the gradient of the potential function are approximated using central finite differences of second order of accuracy for the 4th order modified Hamiltonian
\begin{equation*}
	U_{\mathbf x}^{(1)}\approx \frac{\mathbf U(t_{n+1})-\mathbf U(t_{n-1})}{2\varepsilon}=:\mathbf U^{(1)},
\end{equation*}
where $\varepsilon=h$ for the Verlet, $\varepsilon=h/2$ for two-stage and $\varepsilon=ah$ for  three-stage integrators ($h$ is the integration step size and $a$ the coefficient in \eqref{3S}.
The 6th order modified Hamiltonian, here considered only for the Verlet and two-stage integrators, is calculated using centered differences of fourth order accuracy for the first derivative and second order accuracy for the second and third derivatives
\begin{eqnarray*}
	U_{\mathbf x}^{(1)}&\approx &\frac{\mathbf U(t_{n-2})-8\mathbf U(t_{n-1})+8\mathbf U(t_{n+1})-\mathbf U(t_{n+2})}{12\varepsilon}=:\mathbf U^{(1)} \nonumber \\
	U_{\mathbf x}^{(2)}&\approx &\frac{\mathbf U(t_{n-1})-2\mathbf U(t_n)+\mathbf U(t_{n+1})}{\varepsilon^2} =:\mathbf U^{(2)}\\
	U_{\mathbf x}^{(3)}&\approx&\frac{-\mathbf U(t_{n-2})+2\mathbf U(t_{n-1})-2\mathbf U(t_{n+1})+\mathbf U(t_{n+2})}{2\varepsilon^3}=: \mathbf U^{(3)},\nonumber
\end{eqnarray*}
where $\varepsilon$ depends on the integrator as before.

The final expressions for the  modified Hamiltonians computed in this way are
\begin{eqnarray}
	\tilde{H}^{[4]}(\mathbf x ,\mathbf p)  &=& H(\mathbf x ,\mathbf p) +h k_{21} \mathbf p^T M^{-1}{P_1} + h^2 k_{22} {U_{\mathbf x}(\mathbf x)}^T M^{-1}U_{\mathbf x}(\mathbf x )  \label{modHam4num} \\
	\nonumber \\
	\tilde{H}^{[6]}(\mathbf x ,\mathbf p) &=& \tilde{H}^{[4]}(\mathbf x ,\mathbf p) + h k_{41}\mathbf p^T M^{-1}P_3 + h^2 k_{42} {{U_{\mathbf x }}(\mathbf x)}^TM^{-1}P_2 \label{modHam6num} \\
	&+& h^2 k_{43} P_1^TM^{-1}P_1 +  h^4 k_{44} U_{\mathbf x }(\mathbf x )^TM^{-1} U_{\mathbf x \mathbf x}(\mathbf x ) M^{-1}U_{\mathbf x }(\mathbf x), \nonumber
\end{eqnarray}
where $P_i=\mathbf U^{(i)}\cdot h^i$.
Note that the term with  coefficient $k_{22}$ is calculated exactly,
i.e.\ avoiding finite difference approximation; this improves the approximation
of the modified Hamiltonian when compared to the  strategy originally used in GSHMC. Also note that, compared
to the expressions with analytical derivatives \eqref{modHam4_an} and \eqref{modHam6_an},
in the formulations \eqref{modHam4num}
and \eqref{modHam6num}
the terms involving the coefficients $c_{21}, c_{41}, c_{42}$ and $c_{44}$ are
approximated by $P_i$. The order of accuracy provided
by the modified Hamiltonians  \eqref{modHam4num} and \eqref{modHam6num}, however,
is not affected by these approximations.

\begin{figure}[!ht]
	\centering
	\begin{tikzpicture}
		\begin{groupplot}[group style={group size=2 by 1, horizontal sep=1.7cm, vertical sep=2cm}]
			\nextgroupplot[width=8.1cm, height=5.67cm,
				grid = major,
				xmin=0, xmax=155, xlabel={number of integration steps}, xtick={0,50,100,150}, xticklabels={0,50,100,150},
				ymin=0, ymax=120, ylabel={overhead (\%)}, ytick={0,20,40,60,80,100,120}, yticklabels={0,20,40,60,80,100,120},
				legend pos=north east, legend style={draw=black,fill=white,legend cell align=left}]
	
				\addplot[color=mygrey,dashed,line width=2pt] table[x=L,y=six]{overheads_shadow.dat};
	
				\addplot[color=mygrey,line width=2pt] table[x=L,y=four]{overheads_shadow.dat};
		
			\nextgroupplot[width=8.1cm, height=5.67cm,
				grid = major,
				xmin=0, xmax=155, xlabel={number of integration steps}, xtick={0,50,100,150}, xticklabels={0,50,100,150},
				ymin=0, ymax=200, ylabel={overhead (\%)}, ytick={0,50,100,150,200}, yticklabels={0,50,100,150,200},
				legend pos=north east, legend style={draw=black,fill=white,legend cell align=left}]
	
				\addplot[color=mygrey,line width=2pt] table[x=L,y=four]{overheads_shadow2.dat};
				\addlegendentry{$\tilde{H}^{[4]}$};
	
				\addplot[color=mygrey,dashed,line width=2pt] table[x=L,y=six]{overheads_shadow2.dat};
				\addlegendentry{$\tilde{H}^{[6]}$};
		\end{groupplot}
	\end{tikzpicture}
	\caption{Computational overhead of MMHMC compared to HMC for models with a tridiagonal (left) and a dense  (right) Hessian matrix, using 4th and 6th order modified Hamiltonians with numerical approximation of the time derivatives.}
	\label{Fig:overheads_modHam46N}
\end{figure}
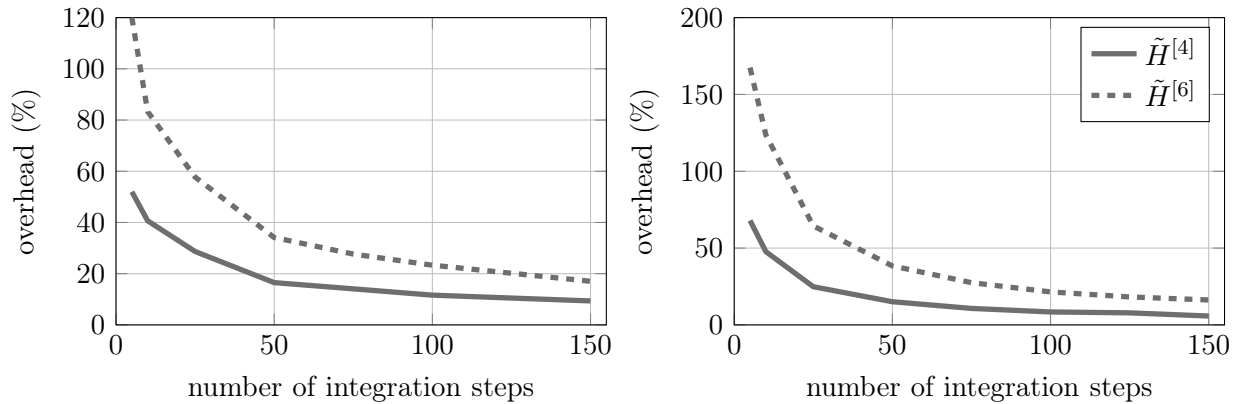

The computational overhead of the MMHMC method \cite{Radivojevic:2017} using the modified Hamiltonians
 \eqref{modHam4num} or  \eqref{modHam6num}, when compared to HMC is
   shown in Figure \ref{Fig:overheads_modHam46N} for models with a tridiagonal (left-hand graph)
    and  dense Hessian matrix (right-hand graph) of the potential, as described for Figure \ref{Fig:overheads_modHam4A}.
Compared to Figure \ref{Fig:overheads_modHam4A}, where all derivatives are calculated analytically,
we note that for models with a sparse Hessian (left-hand graphs), the 4th order modified Hamiltonian
 \eqref{modHam4_an} with analytical derivatives introduces less computational overhead than \eqref{modHam4num}
  with a numerical approximation of the time derivative. This is due to the additional forward and backward
   integration steps, which do not counterbalance the inexpensive Hessian computation.
For models with a dense Hessian matrix (right-hand graphs) we recommend
 always using \eqref{modHam4num}, which significantly reduces the overhead.
The 6th order modified Hamiltonian \eqref{modHam6num} clearly requires additional computational effort,
due to two extra gradient calculations per MC iteration.
In the following sections we show that using  modified Hamiltonians of order higher than 4 can be avoided
 by introducing accurate multi-stage integrators specifically tuned for the MHMC methods.

In summary, we have provided two alternative formulations of the 4th and 6th order modified Hamiltonians corresponding to multi-stage integrators \eqref{2S}--\eqref{3S} with arbitrary coefficients.
For the cases when derivatives of the potential function are available
and inexpensive to compute, the modified Hamiltonians can be calculated
using \eqref{modHam4_an}--\eqref{modHamVerlet}. For problems in which this is not the case,
we provided formulations of modified Hamiltonians which mainly rely on quantities available
 from the simulation. Both approaches can be used with any multi-stage integrator \eqref{2S}--\eqref{3S}, including the Verlet integrator.

In the following section, we choose the coefficients in the families \eqref{2S}--\eqref{3S} so as to obtain methods
specifically aimed at
sampling with modified Hamiltonians.

\section{Multi-stage integrators with optimized coefficients}\label{Sec:Integrators}

Until now, the Verlet/leapfrog integrator has been the integrator of choice for MHMC methods.
In this section, we consider alternative integrators and investigate their competitiveness with the Verlet integrator.

Our focus is on multi-stage integrators belonging to families \eqref{2S}--\eqref{3S}. There are two reasons for an interest in these integrators. One is their potential to achieve, at a given computational cost, higher accuracy than Verlet. Here it is important to emphasize that, in comparisons, different integrators have
 to be applied with the same computational effort, rather than with the same step length;
 an $r$-stage integrator requires $r$ gradient evaluations per time step and to be compared with Verlet
  has to be used with a step length correspondingly longer. 
 More accurate integrations imply higher acceptance rates and  thus better space exploration.
A second possible benefit of the integrators of this class is that, due to the extra accuracy, they may
 avoid the need for computationally expensive, higher order modified Hamiltonians.

Our goal is to derive new multi-stage integrators to be used in the methods which sample with modified Hamiltonians and compare their impact on the performance of such methods with the efficiency of advanced integrators \textcolor{black}{suitable} for HMC \cite{McL95,BCSS14} and the Verlet integrator.

In  MHMC methods, the equations of motion of the Hamiltonian dynamics are the same as in  HMC methods.
However, MHMC are based on different Metropolis tests where the acceptance rate depends on the capability of
the integrator to conserve the value of a modified Hamiltonian. Indeed, the sampling performance of MHMC is
controlled not by the energy error with respect to the true Hamiltonian as in HMC, but by the energy error
with respect to the modified Hamiltonian. Thus, inspired by the ideas of \cite{McL95} and \cite{BCSS14} for
improving HMC performance by minimizing (expected) energy error through the appropriate choice of parameters
of the integrator, in order to enhance the performance of MHMC, we design the new integrators by considering
either the error in the modified Hamiltonians $\tilde{H}^{[l]}$ of order $l$
\begin{equation}\label{modHamError}
	\Delta=\tilde{H}^{[l]}(\Psi_{h,L}(\mathbf x,\mathbf p))-\tilde{H}^{[l]}(\mathbf x,\mathbf p),
\end{equation}
or the expected values of such errors $\mathbb E_{\tilde{\pi}}(\Delta)$ taken with respect to the modified
canonical density $\tilde{\pi}$ given by \eqref{eq:hk}.
Here $\Psi_{h,L}(\mathbf x,\mathbf p)$ is the  $hL$-time map of the integrator.
In order to distinguish the new minimum error and minimum expected error integrators
for sampling with modified (M) Hamiltonians  from the corresponding ones designed for the
HMC method, we use the prefix M-; for instance M-ME will denote minimum error for sampling with modified Hamiltonians.

\subsection{Minimum error (M-ME) integrators}

We first construct the minimum error integrators for the 4th order modified Hamiltonian.

The Taylor expansion of the 4th order modified Hamiltonian after one integration step with the method $\Psi_{h}$ can be written as \citep{SanzSerna94}
\begin{eqnarray*}
	\tilde{H}^{[4]}(\mathbf x{'},\mathbf p{'})&=&\tilde{H}^{[4]}(\Psi_{h}(\mathbf x,\mathbf p))=\exp\big(h\mathcal L_{\tilde{H}}\big)\tilde{H}^{[4]}(\mathbf x,\mathbf p)  \\
	&=& \tilde{H}^{[4]}( \mathbf x,\mathbf p) +h\mathcal L_{\tilde{H}}\tilde{H}^{[4]}(\mathbf x,\mathbf p)+\frac{1}{2}h^2\mathcal L^2_{\tilde{H}}\tilde{H}^{[4]}(\mathbf x,\mathbf p)+\dots ,
\end{eqnarray*}
where $\tilde{H}$ is the modified Hamiltonian \eqref{modifiedHam} and we have used the Lie derivative
 \eqref{LieDer_def}.
The error $\Delta$ in $\tilde{H}^{[4]}$ after one integration step reads
\begin{equation}
	\begin{aligned}
		\Delta(\mathbf x,\mathbf p)=h^5 \big(&\gamma_1 \{A, A, A, A, A, B\}(\mathbf x,\mathbf p)+ \gamma_1\{B, A, A, A, A, B\}(\mathbf x,\mathbf p) \label{Energy_error} \\
		+& \gamma_2\{A,B,A,A,A,B\}(\mathbf x,\mathbf p) + \gamma_2\{B, B, A, A, A, B\}(\mathbf x,\mathbf p)   \\
		+& \gamma_3\{A, B, B, A, A, B\}(\mathbf x,\mathbf p) + \gamma_3\{B, B, B, A, A, B\}(\mathbf x,\mathbf p)   \\
		+& \gamma_4\{A, A, A, B, B, A\}(\mathbf x,\mathbf p) + \gamma_4\{B, A, A, B, B, A\}(\mathbf x,\mathbf p) \big)
+\mathcal{O}(h^6).
	\end{aligned}
\end{equation}

Thus, as $h$ approaches $0$ and the $\mathcal{O}(h^6)$ remainder may be ignored, an error metric for the 4th
order modified Hamiltonian can then be defined by the following function of the integrating coefficients
\begin{equation}\label{minErrMetric}
	E=\sqrt{\gamma_1^2+\gamma_2^2+\gamma_3^2+\gamma_4^2}.
\end{equation}
The  expressions for $\gamma_{1-4}$ derived from \eqref{coeff_c_gamma} are
$$\gamma_1=c_{41}, \hspace{0.5cm} \gamma_2=\frac{1}{3}(c_{44}-c_{42}), \hspace{0.5cm} \gamma_3
=\frac{1}{2}c_{43}, \hspace{0.5cm} \gamma_4=\frac{1}{2}c_{44}.$$
The coefficients $c_{ij}$ are calculated from \eqref{modHam2S} and \eqref{modHam3S} for two- and three-stage integrators, respectively.
For quadratic potential and kinetic energies, corresponding to the problem of sampling from a Gaussian distribution with a covariance matrix $\Sigma$, or  harmonic oscillators, the error \eqref{Energy_error} simplifies to
$$\Delta(\mathbf x,\mathbf p)=h^5(\gamma_4-\gamma_3)(4 \mathbf p^TM^{-1}M^{-1}M^{-1}\Sigma^{-1}\Sigma^{-1}\Sigma^{-1}\mathbf x) +\mathcal{O}(h^6),$$
as the other terms involve third (or higher) order partial derivatives of $A$ and $B$ and are equal to zero.
Therefore, the error metric can be defined as
\begin{equation}\label{minErrMetricG}
	E^Q=|\gamma_4-\gamma_3|.
\end{equation}

In contrast to this approach, \textcolor{black}{the
integrator derived in \citep{McL95} only takes into account the truncation error for the true Hamiltonian,
rather than the error for the modified Hamiltonian after numerical integration, as we do here.}

In order to obtain values for integrating coefficients for the MHMC methods, we minimized the metrics $E$ or $E^{Q}$ on the interval $(0,0.5)$ using \textit{Mathematica}. For the three-stage family of integrators for problems with quadratic potential and kinetic function, i.e.\ for $E^Q$, we take into account the analysis from \cite{Campos:2017}, where the authors provide the condition for which the stability limit is the highest for these kind of integrators. In particular, the integrators that lie on the hyperbola
\begin{equation}\label{eq:hyperbola}
	6ab-2a-b+\frac{1}{2}=0,
\end{equation}
have considerably longer stability limit than others. Therefore, we enforce the relationship between $a$ and $b$ from \eqref{eq:hyperbola} in the minimization of $E^Q$. The resulting values of coefficients for two- and three-stage minimum error integrators for quadratic (M-ME2, M-ME3) and general problems (M-ME2gen, M-ME3gen) are given in Table \ref{Tab:IntSummary}. We note that the difference between the coefficients for the two versions of the two-stage methods is minor.

The error metric definitions \eqref{minErrMetric} and \eqref{minErrMetricG} are based on the assumption that the different iterated brackets that feature  in \eqref{Energy_error} contribute equally to the Hamiltonian error.
While this assumption is reasonable, in problems where information on the relative size of the iterated brackets is available one may modify accordingly the error metric so as to give more weight to the error coefficients with larger iterated bracket; of course that change in metric would entail a change in the coefficient values resulting from the minimization procedure.

\subsection{Minimum expected error integrators (M-BCSS)}

The modified Hamiltonians we consider here are of order 4 or 6. We adopt a strategy similar to the one
proposed in \cite{BCSS14}, namely to consider the problem of a one-dimensional quadratic potential and to find
the parameters of integrators that minimize the expected value of the error for a suitable range of finite
values of $h$; note that this is different from the procedure above based on the behaviour of the algorithms
as $h$ approaches $0$. In our case, the error \eqref{modHamError}, resulting from numerical integration, is in
terms of the modified Hamiltonian and the expected value is taken with respect to the modified canonical
density $\tilde{\pi}$.

As in the case when considering the error in the true Hamiltonian, one may prove that
 the expected error in the modified Hamiltonian $\mathbb E_{\tilde{\pi}}(\Delta)$ is also positive.
The objective is, therefore, to find a function $\rho(h,\boldsymbol{\xi})$ that bounds $\mathbb E_{\tilde{\pi}}(\Delta)$, i.e.
$$0 \leq \mathbb E_{\tilde{\pi}}(\Delta) \leq \frac{1}{\beta} \rho(h,\boldsymbol{\xi}).$$
Here $\boldsymbol{\xi}$ is a parameter vector, i.e.\ $\boldsymbol{\xi}=\{b \}$ for two-stage integrators,  $\boldsymbol{\xi}=\{a,b \}$ for three-stage integrators, and $h$ is a dimensionless step size.
We omit here the derivation of $\rho(h,\boldsymbol{\xi})$ as it can be found in \citep{Akhmatskaya:2017}
and provide the expression for $\rho$ for the families of two- and three-stage integrators when sampling
with 4th and 6th order modified Hamiltonians. This expression if of the form
\begin{equation}\label{rho_MHMC}
	{\rho}(h,\boldsymbol\xi)=\frac{\Big(SB_h+C_h\Big)^2}{2S(1-A^2_h)}.
\end{equation}
The symbols in the right-hand side are as follows. For the 4th order modified Hamiltonian
$$S=\frac{1+2h^2c_{22}}{1+2h^2c_{21}}$$
 and for the 6th order modified Hamiltonian
$$S=\frac{1+2h^2c_{22}+2h^4c_{43}}{1+2h^2c_{21}+2h^4c_{44}},$$
where coefficients $c_{ij}$ depend on the integrator and were derived in Section \ref{Sec:Mod_Ham}.
The $h$-dependent quantities $A_h,B_h,C_h$ come from the matrix $\tilde{M}_h$ that advances the numerical solution
over a single time step in the integration of the harmonic oscillator:
\begin{eqnarray*}
\begin{bmatrix}
x_{n+1}\\
p_{n+1}
\end{bmatrix} = \tilde{M}_h
\begin{bmatrix}
x_{n}\\
p_{n}
\end{bmatrix}, \quad
\tilde{M}_h= \begin{bmatrix}
			A_h & B_h \\
			C_h & A_h
			\end{bmatrix}.
\end{eqnarray*}
For two-stage integrators, the matrix $\tilde{M}_h$ has the expression
\begin{equation*}
\tilde{M}_h= B\left(b\right)\cdot A\left(\frac1 2\right)\cdot B\left(1-2b\right)\cdot A\left(\frac1 2\right)\cdot B\left(b\right),
\end{equation*}
where
$$A(a)= \begin{bmatrix}
			1 & ah \\
			0 & 1
			\end{bmatrix}, \; B(b)= \begin{bmatrix}
			1 & 0 \\
			-bh & 1
			\end{bmatrix}$$
correspond to mappings  $\varphi^A_h$ and $\varphi^B_h$, respectively. The resulting elements of $\tilde{M}_h$  are
then
\begin{eqnarray*}
A_h&=& \frac{h^4}{4}b(1-2b)-\frac{h^2}{2}+1,  \\
B_h&=& -\frac{h^3}{4}(1-2b)+h,\\
C_h&=& -\frac{h^5}{4}b^2(1-2b)+h^3b(1-b)-h.
\end{eqnarray*}
Similarly, for the three-stage family we compute
$$\tilde{M}_h= B(b)\cdot A(a)\cdot B(\frac 1 2-b)\cdot A(1 -2a)\cdot B(\frac 1 2-b)\cdot A(a)\cdot B(b) $$
and obtain
\begin{eqnarray*}
A_h&=& \frac{h^6}{4} a^2 (2 a-1) (1 - 2 b)^2 b + \frac{h^4}{4}a \left(1 - 4 b^2 - a (1 - 4 b)\right) -\frac{h^2}{2}+1  \\
B_h&=& \frac{h^5}{4}a^2 (1 - 2 a) (1 - 2 b)^2  - h^3 a (1-a)(1-2 b)  +h \\
C_h&=& \frac{h^7}{4} a^2 (1 - 2 a) (1 - 2 b)^2 b^2+
	 \frac{ h^5}{2} a ( 2 a (1 - b)-1) b (1 - 2 b)+  \\
	&& \frac{h^3}{4}  \left(1 - 2 a (1 - 2 b)^2\right)  -h.
\end{eqnarray*}

Note that the true Hamiltonian can be recovered by setting the coefficients $c_{ij}$ to zero. Doing so,
we obtain exactly the same function derived in \cite{BCSS14}
\begin{equation}\label{rho_HMC}
\rho_{\text{HMC}}(h,\boldsymbol\xi)=\frac{(B_h+C_h)^2}{2(1-A^2_h)}.
\end{equation}

The coefficients $\boldsymbol\xi$ can be found by minimizing the function
\begin{equation}\label{norm_rho}
	\| \rho \|_{(\bar{h})}=\max_{0<h<\bar{h}} \rho(h,\boldsymbol\xi),
\end{equation}
where $\bar h$ is equal to the number of stages in the integrator \citep{BCSS14}. Thus we obtain the parameter $b=0.238016$ for the two-stage M-BCSS2 integrator derived for sampling with the MHMC methods. We note the difference in value for the coefficient of the original two-stage BCSS2 integrator, $b=0.21178$, introduced for HMC and obtained by minimizing the function \eqref{norm_rho} using \eqref{rho_HMC}.

Using again the stability analysis from \cite{Campos:2017} for three-stage integrators, namely enforcing the
condition \eqref{eq:hyperbola}, we obtain the coefficients $b=0.1441153, a=(1-2b)/4(1-3b)$ for the M-BCSS3
integrator for sampling with MHMC.

\begin{figure}[!ht]
	\centering
	\begin{tikzpicture}
	\begin{groupplot}[group style={group size=2 by 1, horizontal sep=2cm, vertical sep=2cm}]
	\nextgroupplot[width=8cm, height=5.6cm,
		xmin=0, xmax=5, xlabel={$\bar{h}$}, xtick={0,1,2,3,4,5}, xticklabels={0,1,2,3,4,5},
		ymin=0, ymax=25, ylabel={$\max_{0 < h < \bar{h}}{\rho(h)}$}, ytick={0,5,10,15,20,25}, yticklabels={0,0.05,0.1,0.15,0.2,0.25},
		legend style={at={(0,0)},draw=black,fill=white,legend cell align=left, anchor=south},legend columns=3,legend to name=legbound]
	
		\addplot [restrict y to domain=0:30] [color=mygrey,line width=1pt,]
		table[x=x,y=V]{rho_small100.dat};
		\addlegendentry{V};
	
		\addplot [restrict y to domain=0:30] [color=myblue,dashed,line width=1pt,]
		table[x expr={\thisrow{x}*3/2},y=ME]{rho_small100.dat};
		\addlegendentry{ME};

		\addplot [restrict y to domain=0:30] [color=myred,dashed,line width=1pt,]
		table[x expr={\thisrow{x}*3/2},y=BCSS2]{rho_small100.dat};
		\addlegendentry{BCSS2};

		\addplot [restrict y to domain=0:30] [color=myred,solid,line width=1pt,]
		table[x=x,y=BCSS3]{rho_small100.dat};
		\addlegendentry{BCSS3};

		\addplot [restrict y to domain=0:30] [color=mygrey,line width=2pt,]
		table[x=x,y=MV]{rho_small100.dat};
		\addlegendentry{M-V};

		\addplot [restrict y to domain=0:30] [color=myblue,dashed,line width=2pt,]
		table[x expr={\thisrow{x}*3/2},y=MME2]{rho_small100.dat};
		\addlegendentry{M-ME2};

		\addplot [restrict y to domain=0:30] [color=myblue,solid,line width=2pt,]
		table[x=x,y=MME3]{rho_small100.dat};
		\addlegendentry{M-ME3};

		\addplot [restrict y to domain=0:30] [color=myred,dashed,line width=2pt,]
		table[x expr={\thisrow{x}*3/2},y=MBCSS2]{rho_small100.dat};
		\addlegendentry{M-BCSS2};

		\addplot [restrict y to domain=0:30] [color=myred,solid,line width=2pt,]
		table[x=x,y=MBCSS3]{rho_small100.dat};
		\addlegendentry{M-BCSS3};
		
	\coordinate (c1) at (rel axis cs:0,1);
		
	\nextgroupplot[width=8cm, height=5.6cm,
		xmode=log, xmin=1e-2, xmax=5, xlabel={$\bar{h}$}, xtick={1e-2,1e-1,1e0}, xticklabels={$10^{-2}$,$10^{-1}$,$10^0$},
		ymode=log, ymin=1e-25, ymax=1e2, ylabel={$\max_{0 < h < \bar{h}}{\rho(h)}$}, ytick={1e-25,1e-20,1e-15,1e-10,1e-5,1e0}, yticklabels={$10^{-25}$,$10^{-20}$,$10^{-15}$,$10^{-10}$,$10^{-5}$,$10^0$},]
		
		\addplot [color=mygrey,line width=1pt,]
		table[x=x,y=V]{rho_small100.dat};

		\addplot [color=myblue,dashed,line width=1pt,]
		table[x expr={\thisrow{x}*3/2},y=ME]{rho_small100.dat};

		\addplot [color=myred,dashed,line width=1pt,]
		table[x expr={\thisrow{x}*3/2},y=BCSS2]{rho_small100.dat};

		\addplot [color=myred,solid,line width=1pt,]
		table[x=x,y=BCSS3]{rho_small100.dat};

		\addplot [color=mygrey,line width=2pt,]
		table[x=x,y=MV]{rho_small100.dat};

		\addplot [color=myblue,dashed,line width=2pt,]
		table[x expr={\thisrow{x}*3/2},y=MME2]{rho_small100.dat};

		\addplot [color=myblue,solid,line width=2pt,]
		table[x=x,y=MME3]{rho_small100.dat};

		\addplot [color=myred,dashed,line width=2pt,]
		table[x expr={\thisrow{x}*3/2},y=MBCSS2]{rho_small100.dat};

		\addplot [color=myred,solid,line width=2pt,]
		table[x=x,y=MBCSS3]{rho_small100.dat};
		
	\coordinate (c2) at (rel axis cs:1,1);
	\end{groupplot}
	\coordinate (c3) at ($(c1)!.5!(c2)$);
	\node[below] at (c3 |- current bounding box.south) {\pgfplotslegendfromname{legbound}};
	\end{tikzpicture}
	\caption{Upper bound for the expected energy error for the two- and three-stage  \mbox{(M-)BCSS},  (M-)ME and Verlet integrators for sampling with the true Hamiltonian (dashed) and 4th order modified Hamiltonian (solid). Right-hand graph shows the same functions on a logarithmic scale.}\label{Rho_func}
\end{figure}
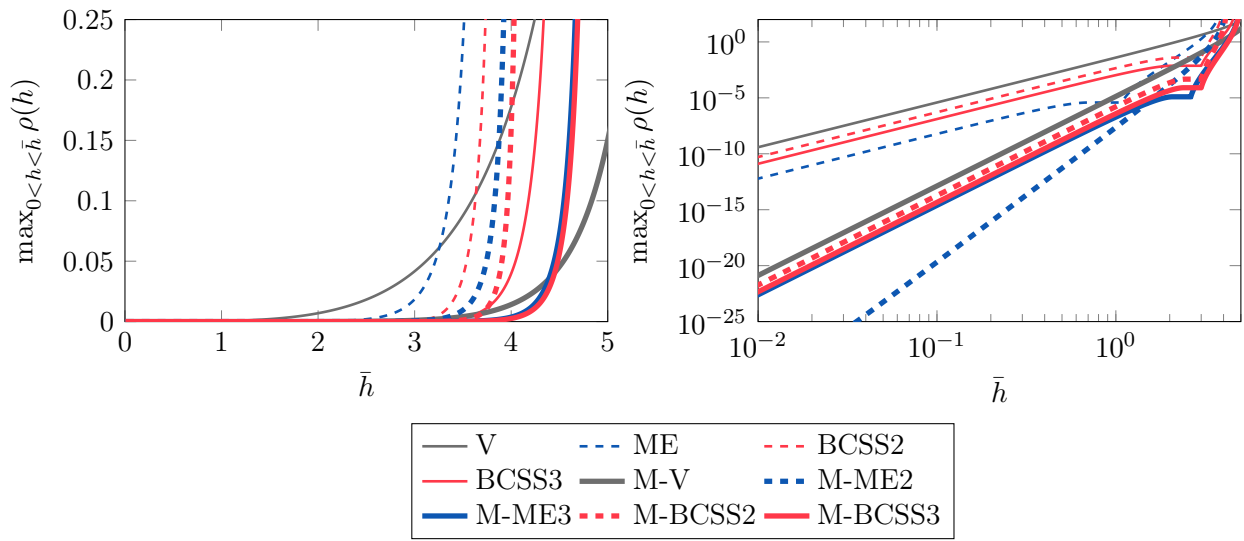

In Figure \ref{Rho_func} $\| \rho_{\text{HMC}} \|_{(\bar{h})}$ from \eqref{norm_rho} and \eqref{rho_HMC} is plotted as a function of the maximal step size $\bar h$
(here normalized to the three-stage schemes, i.e.\ $\bar{h}_{r\text{-stage}}=r\cdot\bar{h}/3, r=1,2,3$)
 for the two- and three-stage integrators for the HMC method (dashed lines), and the corresponding function
  $\| \rho \|_{(\bar{h})}$ from \eqref{rho_MHMC} and \eqref{norm_rho} for the two- and three-stage integrators, derived in this section for sampling with MHMC (solid lines).
The functions $\| \rho_{\text{HMC}} \|_{(\bar{h})}$ and $\| \rho \|_{(\bar{h})}$ for the Verlet integrator are also plotted.
We note that the upper bound of the expected error in Hamiltonian, or modified Hamiltonian, and thus the error of the method, is lower for integrators derived for MHMC than in the case of the  HMC specific integrators, which confirms a better conservation of modified Hamiltonians than true Hamiltonians  by symplectic integrators.
As follows from Figure \ref{Rho_func}, the multi-stage integrators derived for HMC and MHMC provide better accuracy than Verlet for step sizes less or equal to a half stability limit of Verlet, i.e.\ $\bar h=3$, with three-stage integrators being superior to two-stage class.
Please notice that $\bar h$ in Figure \ref{Rho_func} refers to a step size for a three-stage integrator. If Verlet is viewed as a single stage integrator, its half stability limit corresponds to $\bar h=1$.
The integrators derived for MHMC guarantee a better accuracy than other
integrators for $\bar h$ even bigger than $3$, which implies their efficiency for longer step sizes compared
with Verlet and multi-stage integrators for HMC. A logarithmic scale version of the left-hand graph, shown in
the right-hand graph, gives a better insight into the  behavior of the functions.

It is important to note that the Verlet integrator has the largest stability interval\footnote{The stability interval of an integrator is defined as the largest interval $(0,h_{\max})$ such that the integrator is stable for all $h \in(0,h_{\max})$.} among other multi-stage integrators, and due to this, care should be taken of the choice of the step size. \textcolor{black}{This fact has been considered in previous studies \cite{FPASS16,Bou-Rabee:2018,Akhmatskaya:2017}.}
The stability intervals computed for the considered model problem and for each of the examined integrators are given in Table \ref{Tab:IntSummary} in terms of the three-stage family. We note that the trends of the stability limit $h_{\max}$ for each integrator are in agreement with the corresponding upper bound functions.
Nevertheless, as Figure \ref{Rho_func} suggests, the accuracy is degrading with $\bar h$ approaching the stability limit.
It is the characteristics of the sampling problem (such as the dimension of the system,  number of observations,  nature of the underlying model)  that determine the optimal step size and therefore the integrator which would provide the best performance.

 \begin{table}[!h]
	\centering
	\begin{tabular}{cccccc}
		Integrator & Application & N.\ of stages & Coefficients & $h_{\max}$ & Ref. \\ \hline
		Verlet & HMC, MHMC & 1 & -- & 6.000 & \cite{DKPR87,Neal10,IH04,Akhmatskaya08} \\ \hline
		BCSS2 & HMC & 2 & $b=0.21178\textcolor{black}{1}$ & 3.951 & \cite{BCSS14} \\ \hline
		M-BCSS2 & MHMC & 2 & $b=0.238016$ & 4.144 & this work \\ \hline
		ME & HMC & 2 & $b=0.193183$ & 3.830 & \cite{McL95} \\ \hline
		M-ME2 & MHMC & 2 & $b=0.230907$ & 4.089 & this work \\ \hline
		M-ME2gen & MHMC & 2 & $b=0.23061\textcolor{black}{0}$ & 4.087 & this work \\ \hline
		\multirow{2}{*}{BCSS3} & \multirow{2}{*}{HMC} & \multirow{2}{*}{3} & $a=(1-2b)/4(1-3b)$ & \multirow{2}{*}{4.662} & \multirow{2}{*}{\cite{BCSS14,Campos:2017}} \\
		&&& $b= 0.11888\textcolor{black}{0}$ && \\ \hline
		\multirow{2}{*}{M-BCSS3} & \multirow{2}{*}{MHMC} & \multirow{2}{*}{3} & $a=(1-2b)/4(1-3b)$ & \multirow{2}{*}{4.902} & \multirow{2}{*}{this work} \\
		&&& $b=0.144115$ && \\ \hline
		\multirow{2}{*}{M-ME3} & \multirow{2}{*}{MHMC} & \multirow{2}{*}{3} & $a=(1-2b)/4(1-3b)$ & \multirow{2}{*}{4.887} & \multirow{2}{*}{this work} \\
		&&& $b=0.142757$ && \\ \hline
		\multirow{2}{*}{M-ME3gen} & \multirow{2}{*}{MHMC} & \multirow{2}{*}{3} & $a=0.355423$ & \multirow{2}{*}{\textcolor{black}{2.986}} & \multirow{2}{*}{this work} \\
		&&& $b=0.184569$ && \\ \hline
	\end{tabular}
	\caption{The splitting integrators for sampling with the true or 4th order modified Hamiltonians developed or tested in this study.
	Stability limit $h_{\max}$
	 is presented in terms of the three-stage family.}
	\label{Tab:IntSummary}
\end{table}

\section{Numerical experiments}\label{Sec:Experiments}

In this section we examine the performance of the novel schemes on two benchmark models and compare them against Verlet and  integrators suggested in the literature to be used within HMC algorithms.

\subsection{Algorithms and performance metrics used}

For the numerical experiments we employ the GSHMC \cite{Akhmatskaya08} and MMHMC \cite{Radivojevic:2017} algorithms, introduced for sampling in molecular simulation and computational statistics problems, respectively.
The main feature of these two algorithms that distinguishes them from other MHMC methods is the momentum update step, called Partial Momentum Monte Carlo (PMMC).
 In this step, the partial momentum refreshment is combined with a modified Metropolis test.
 Namely, for the current momentum $\mathbf {p}$ and a noise vector $\mathbf {u}\sim \mathcal N(0,\beta^{-1}M)$  a proposal in GSHMC  is defined as
\begin{equation*}
\begin{aligned}
	\mathbf p^* &=& \cos(\phi)\mathbf p+\sin(\phi)\mathbf u  \\
	\mathbf u^*&=& -\sin(\phi)\mathbf p +\cos(\phi)\mathbf u,
\end{aligned}
\end{equation*}
where the parameter $\phi \in (0,\pi/2]$ controls the amount of noise introduced in every iteration.
In MMHMC, the momentum update is defined through the noise parameter $\varphi \in (0,1]$ ($\varphi=\sin^2(\phi)$) as
\begin{equation*}
\begin{aligned}
	\mathbf p^* &=& \sqrt{1-\varphi}\mathbf{p} + \sqrt{\varphi}\mathbf {u} \\
	\mathbf u^* &=& -\sqrt{\varphi}\mathbf{p} + \sqrt{1-\varphi}\mathbf {u}.
\end{aligned}
\end{equation*}
In the {modified Metropolis test}, \textcolor{black}{in both GSHMC and MMHMC,} the proposal $(\mathbf p^*,\mathbf u^*)$ is accepted according to
\begin{equation*}
(\bar{\mathbf p},\bar{\mathbf u})=\left\{\begin{array}{l l}
                (\mathbf p^*,\mathbf u^*) & \mbox{with probability } \mathcal{P}\\
                 (\mathbf p,\mathbf u) & \mbox{otherwise,}
                \end{array}\right.
\end{equation*}
where
\begin{equation*}
	\mathcal{P}=\min\left\{1,{\frac{\exp\big(-(\tilde{H}(\mathbf x,\mathbf p^*)+{\frac{1}{2}}(\mathbf u^*)^{T}M^{-1}\mathbf u^*)\big)}{\exp\big(-(\tilde{H}(\mathbf x,\mathbf p)+{\frac{1}{2}}\mathbf u^{T}M^{-1}\mathbf u)\big)}}\right\}.
\end{equation*}
This step can be considered as a standard HMC method in which the vector $\mathbf{x}$ is fixed, the vector
$\mathbf p$ plays a role of the ``position'' and the noise vector $\mathbf u$ becomes the ``conjugate
momenta''. The extended Hamiltonian
\begin{equation*}
	\hat{H}(\mathbf{x},\mathbf p,\mathbf u)=\tilde{H}(\mathbf{x},\mathbf p)+{\frac{1}{2}}\mathbf u^{\intercal}M^{-1}\mathbf u
\end{equation*}
defines the extended reference density $\hat{\pi}(\mathbf{x},\mathbf p,\mathbf u)\propto {\exp(-\beta\hat{H}(\mathbf{x},\mathbf p,\mathbf u))}$.

Note that in the case of the 4th order modified Hamiltonian with analytical derivatives of the potential
available, there is no need for additional integration steps forward and backward nor calculation of gradients
within the PMMC step (see details in \cite{Radivojevic:2017}). Otherwise, additional integration steps need to
be performed in order to evaluate the modified Hamiltonian $\tilde{H}(\mathbf x,\mathbf p^*)$ with the
proposed momenta $\mathbf p^*$. The noise vectors $\mathbf u$ and $\bar{\mathbf u}$ are discarded once the
momentum has been updated.

The GSHMC and MMHMC methods have been implemented in the in-house software packages MultiHMC-GROMACS \cite{Akhmatskaya:2017} and HaiCS (Hamiltonians in Computational Statistics) \cite{Radivojevic:2017}, respectively, both written in C and targeted to computers running UNIX certified operating systems.

\medskip

\textcolor{black}{The following indicators have been monitored:}
\begin{itemize}
	\item Acceptance rate (AR);
	\item Effective Sample Size (ESS)---indicating the number of effectively uncorrelated samples out of $N$ collected samples;
	\item Monte Carlo Standard Error (MCSE)---reflects how much error is in the estimate  \eqref{eq:Reweighting} due to the use of a Monte Carlo method. It is related to
 ESS as $MCSE=\sqrt{{\hat{\sigma}^2}/{ESS}}$, where $\hat{\sigma}^2$ is the sample variance;
	\item Total distance from the mean, defined as $\|\mathbf x-\boldsymbol{\mu}\|=\sum_{d=1}^{D}|\hat{x}_d -\mu_d|$ for the mean $\boldsymbol{\mu}$.
\end{itemize}

The ESS and MCSE metrics are calculated as proposed in \cite{Radivojevic:2017} for  methods that produce
samples that are both correlated and weighted.  
\textcolor{black}{The numerical results below present absolute and relative values of these metrics; larger values signal better sampling.
Relative values are
values normalized with respect to Verlet, so that a relative value above 1 means an improvement over Verlet.}

To make the comparison among the schemes fair, the following issues have been taken into account while
producing the numerical results. The step size $h$ and the number of integration steps $L$ were adjusted to
the number of stages in the integrator in such a way that the computational cost is equal for
all tested integrators, i.e.\ for an $r$-stage integrator we set $h_{r\text{-stage}}=rh_{\text{Verlet}}$ and
$L_{r\text{-stage}}=L_{\text{Verlet}}/r$. In all plots of numerical results, values of step sizes correspond
to Verlet and imply step sizes $r$ times bigger for $r$-stage integrators. Each individual test has been
repeated 10 times; the results reported  are obtained by averaging over the 10 runs to reduce statistical
errors.

\subsection{Benchmarks}

Since we are expecting improvements over the Verlet integrator in problems where harmonic oscillations,
 or quadratic terms, are dominating, we choose to test the novel multi-stage integrators on  benchmarks falling
 in this category.
 One obvious choice is a molecular simulation application, namely a benchmark describing a
 realistic coarse-grained system that corresponds to a spider venom toxin in a bilayer. In the following,
  we shall refer to this systems as \textit{toxin}. Gating-modifier toxins, such as spider venom toxin, are of interest as tools for probing channel-structure functions and the biophysical mechanisms of toxin blockade. The system has been previously studied in \cite{Swartz:1995,Li-Smerin:1998,Lee:2004} and served as a benchmark in \cite{Wee08,Akhmatskaya:2011, FPERA14,FPASS16,Akhmatskaya:2017}.
The other benchmark selected for this study consists of sampling from multivariate Gaussian distributions.
This is a model highly relevant to computational statistics where many important realistic
problems involve distributions that are almost Gaussian or mixtures of Gaussians, see \cite{Tripuraneni:2017,Nishimura:2017,Lan:2016, Rasmussen:1996} among many others.

We sample these two benchmarks with the GSHMC and MMHMC methods, respectively.

\subsubsection{Toxin}
Toxin is a coarse-grained system describing a VSTx1 toxin in a POPC bilayer \cite{Jung2005}. Four heavy particles are represented on average as one sphere \cite{Wallace2007,Shih2006}, which produces a total number of 7810 particles. In the simulations performed, Coulomb and van der Waals interactions were solved using the shift algorithm. Both potential energies were shifted to 0~kJ\,mol$^{-1}$ at the radius of 1.2~nm. Periodic boundary conditions were considered in all directions. The target temperature was chosen to be 310~K. No constraints were defined for this system. The tests were run over a range of time steps $h$. The total length of all simulations was 20~ns, which was sufficient for equilibration of the system for those choices of time steps that provided a stable integration. Different lengths of MD trajectories $L$ were also tested. For the sake of clarity, in all tests presented here the length of MD trajectories was fixed to 4000 steps for Verlet and scaled correspondingly for two- and three-stage integrators. These values were found to be good choices for GSHMC with different integration schemes \cite{Akhmatskaya:2017}. The angle $\phi$ used for the momentum refreshment was set to 0.2 and the modified Hamiltonian \eqref{modHam4num} was used for all tests.

\begin{figure}[!ht]
	\centering
	\begin{tikzpicture}
		\begin{groupplot}[group style={group size=2 by 1, horizontal sep=1.7cm, vertical sep=2cm}]
			\nextgroupplot[width=8.1cm, height=5.67cm,
				xmin=9.5, xmax=25.5, xlabel={$h$ (fs)}, xtick={10,15,20,25}, xticklabels={10,15,20,25},
				ymin=40, ymax=100, ytick={25,50,75,100}, yticklabels={25,50,75,100}, ylabel={$AR$ (\%)},
				legend style={at={(0,0)},draw=black,fill=white,legend cell align=left,anchor=south},legend columns=4,legend to name=legacceptance]
			
				\addplot[color=mygrey,mark=*,mark options={solid},line width=2pt] table[x=x,y=V]{ar.dat};
				\addlegendentry{V};
			
				\addplot[color=myblue,dashed,mark=*,mark options={fill=white,solid},line width=1pt] table[x=x,y=ME]{ar.dat};
				\addlegendentry{ME};
			
				\addplot[color=myred,dashed,mark=*,mark options={fill=white,solid},line width=1pt] table[x=x,y=BCSS2]{ar.dat};
				\addlegendentry{BCSS2};
			
				\addplot[color=myred,solid,mark=*,line width=1pt] table[x=x,y=BCSS3]{ar.dat};
				\addlegendentry{BCSS3};
			
				\addplot[color=myblue,dashed,mark=*,mark options={fill=white,solid},line width=2pt] table[x=x,y=MME2]{ar.dat};
				\addlegendentry{M-ME2};
			
				\addplot[color=myblue,solid,mark=*,line width=2pt] table[x=x,y=MME3]{ar.dat};
				\addlegendentry{M-ME3};
			
				\addplot[color=myred,dashed,mark=*,mark options={fill=white,solid},line width=2pt] table[x=x,y=MBCSS2]{ar.dat};
				\addlegendentry{M-BCSS2};
			
				\addplot[color=myred,solid,mark=*,line width=2pt] table[x=x,y=MBCSS3]{ar.dat};
				\addlegendentry{M-BCSS3};
				
			\coordinate (c1) at (rel axis cs:0,1);

			\nextgroupplot[width=8.1cm, height=5.67cm,
				xmin=9.5, xmax=25.5, xlabel={$h$ (fs)}, xtick={10,15,20,25}, xticklabels={10,15,20,25},
				ymin=305, ymax=315, ytick={305,310,315}, yticklabels={305,310,315}, ylabel={$T$ (K)}]
			
				\addplot[color=mygrey,mark=*,mark options={solid},line width=2pt] table[x=x,y=V]{t.dat};
			
				\addplot[color=myblue,dashed,mark=*,mark options={fill=white, solid},line width=1pt] table[x=x,y=ME]{t.dat};
			
				\addplot[color=myred,dashed,mark=*,mark options={fill=white,solid},line width=1pt] table[x=x,y=BCSS2]{t.dat};
			
				\addplot[color=myred,solid,mark=*,line width=1pt] table[x=x,y=BCSS3]{t.dat};
			
				\addplot[color=myblue,dashed,mark=*,mark options={fill=white,solid},line width=2pt] table[x=x,y=MME2]{t.dat};
			
				\addplot[color=myblue,solid,mark=*,line width=2pt] table[x=x,y=MME3]{t.dat};
			
				\addplot[color=myred,dashed,mark=*,mark options={fill=white,solid},line width=2pt] table[x=x,y=MBCSS2]{t.dat};
			
				\addplot[color=myred,solid,mark=*,line width=2pt] table[x=x,y=MBCSS3]{t.dat};
				
			\coordinate (c2) at (rel axis cs:1,1);
		\end{groupplot}
		\coordinate (c3) at ($(c1)!.5!(c2)$);
		\node[below] at (c3 |- current bounding box.south) {\pgfplotslegendfromname{legacceptance}};
	\end{tikzpicture}
	\caption{Toxin. Acceptance rates (left) and temperatures (right) as functions of the time step $h$. Comparison of the two-stage \mbox{(M-)BCSS2}, (M-)ME(2), three-stage (M-)BCSS3, (M-)ME(3), and Verlet integrators.}\label{ARToxin}
\end{figure}
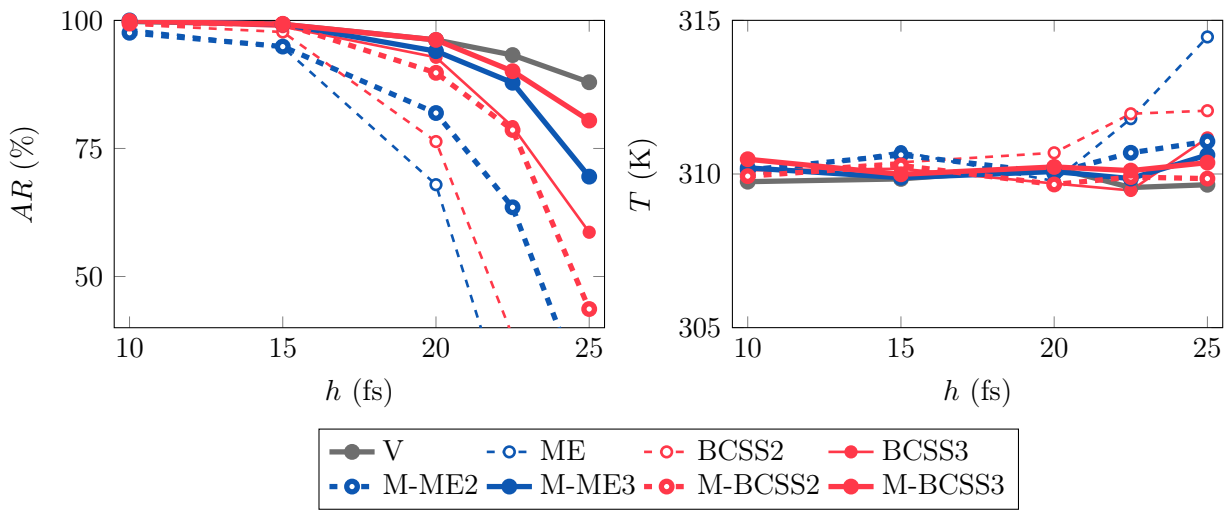

\textcolor{black}{Before we compare the sampling performance of the different methods,} we start by measuring
the acceptance rates in the GSHMC simulations with different multi-stage integration schemes. A fundamental
feature of the GSHMC method is that it maintains very high acceptance rates. \textcolor{black}{We are
interested in this fact since it is known that it has a direct effect in the improvement of the sampling
efficiency \cite{Akhmatskaya08}.} \textcolor{black}{The high acceptance rates are} confirmed in
Figure~\ref{ARToxin} (left), where the effect of various multi-stage integrators and the standard Verlet on
the acceptance rates in GSHMC simulations is
 presented. For small time steps, all integrators provide high acceptance rates, but the situation changes
  as the time step increases and the shorter stability intervals of the different multi-stage methods
  (cf.\ Table \ref{Tab:IntSummary}) result in acceptance rates  below those achieved with Verlet.
We observe that the integrators derived specifically for sampling with modified Hamiltonians in general
 show better acceptance rates than their non-modified counterparts.
Moreover, \textcolor{black}{for multi-stage schemes,} the M-BCSS3 integrator provides the
 best conservation of the modified Hamiltonians and thus the highest acceptance rates.
\textcolor{black}{For time steps \(\leq 20\) fs the acceptance rate of M-BCSS3 is essentially
the same as that of Verlet; however for very long time steps Verlet, as it is well known},
provides the highest acceptance rate due to its better stability. The trends presented in
Figure~\ref{ARToxin} (left) are in a good agreement with the theoretical predictions shown in Figure~\ref{Rho_func}.
 \textcolor{black}{This discussion, however, does not necessarily imply that Verlet is the integrator
  which would guarantee the best sampling performance of GSHMC for toxin. On the contrary, as we show below
  and due to the larger inaccuracies, the sampling performance of GSHMC when used with any tested integrator
   degrades as the time step approaches the  stability limit of the integrator. Thus, the best performance
   is observed for moderate values of time steps.}

\textcolor{black}{The averages of the simulated temperatures $T$ provide a check of the samples obtained.}
 From Figure~\ref{ARToxin} (right)
 it can be observed that all the methods are able to produce the desired averaged temperature with the exception
  of the two-stage methods derived for HMC, which, for the biggest time steps, yield unrealistically high
   temperatures as a result of the very low acceptance rates observed during the simulations in these cases.

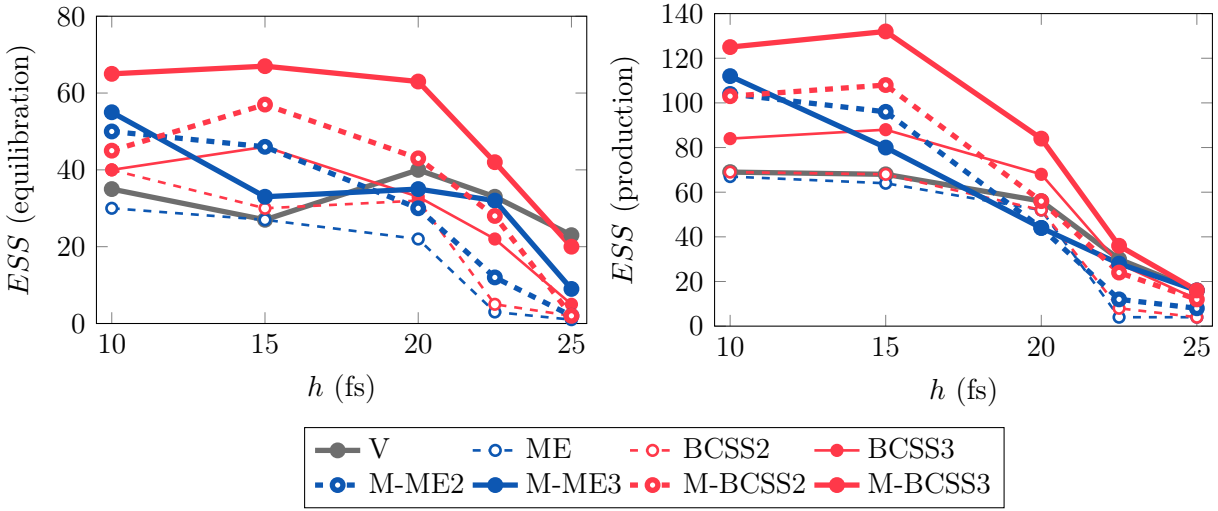
\begin{figure}[!ht]
	\centering
	\begin{tikzpicture}
		\begin{groupplot}[group style={group size=2 by 1, horizontal sep=1.7cm, vertical sep=2cm}]
			\nextgroupplot[width=8.1cm, height=5.67cm,
				xmin=9.5, xmax=25.5, xlabel={$h$ (fs)}, xtick={10,15,20,25}, xticklabels={10,15,20,25},
				ymin=0, ymax=80, ytick={0,20,40,60,80}, yticklabels={0,20,40,60,80}, ylabel={$ESS$ (equilibration)},
				legend style={at={(0,0)},draw=black,fill=white,legend cell align=left, anchor=south},legend columns=4,legend to name=legess]
								
				\addplot[color=mygrey,mark=*,line width=2pt] table[x=x,y=V]{essAbs};
				\addlegendentry{V};
								
				\addplot[color=myblue,dashed,mark=*,mark options={fill=white,solid},line width=1pt] table[x=x,y=ME]{essAbs};
				\addlegendentry{ME};
							
				\addplot[color=myred,dashed,mark=*,mark options={fill=white,solid},line width=1pt] table[x=x,y=BCSS2]{essAbs};
				\addlegendentry{BCSS2};
								
				\addplot[color=myred,solid,mark=*,line width=1pt] table[x=x,y=BCSS3]{essAbs};
				\addlegendentry{BCSS3};
							
				\addplot[color=myblue,dashed,mark=*,mark options={fill=white,solid},line width=2pt] table[x=x,y=MME2]{essAbs};
				\addlegendentry{M-ME2};
								
				\addplot[color=myblue,solid,mark=*,line width=2pt] table[x=x,y=MME3]{essAbs};
				\addlegendentry{M-ME3};
								
				\addplot[color=myred,dashed,mark=*,mark options={fill=white,solid},line width=2pt] table[x=x,y=MBCSS2]{essAbs};
				\addlegendentry{M-BCSS2};
								
				\addplot[color=myred,solid,mark=*,line width=2pt] table[x=x,y=MBCSS3]{essAbs};
				\addlegendentry{M-BCSS3};
				
			\coordinate (c1) at (rel axis cs:0,1);

			\nextgroupplot[width=8.2cm, height=5.74cm,
				xmin=9.5, xmax=25.5, xlabel={$h$ (fs)}, xtick={10,15,20,25}, xticklabels={10,15,20,25},
				ymin=0, ymax=140, ytick={0,20,40,60,80,100,120,140}, yticklabels={0,20,40,60,80,100,120,140}, ylabel={$ESS$ (production)}]
								
				\addplot [color=mygrey, mark=*, line width=2pt] table[x=x,y=V]{essAbsProd};
								
				\addplot [color=myblue,dashed,mark=*,mark options={fill=white,solid},line width=1pt] table[x=x,y=ME]{essAbsProd};
							
				\addplot [color=myred,dashed,mark=*,mark options={fill=white,solid},line width=1pt] table[x=x,y=BCSS2]{essAbsProd};
								
				\addplot [color=myred,solid,mark=*,line width=1pt] table[x=x,y=BCSS3]{essAbsProd};
							
				\addplot [color=myblue,dashed,mark=*,mark options={fill=white,solid},line width=2pt] table[x=x,y=MME2]{essAbsProd};
								
				\addplot [color=myblue,solid,mark=*,line width=2pt] table[x=x,y=MME3]{essAbsProd};
								
				\addplot [color=myred,dashed,mark=*,mark options={fill=white,solid},line width=2pt] table[x=x,y=MBCSS2]{essAbsProd};
								
				\addplot [color=myred,solid,mark=*,line width=2pt] table[x=x,y=MBCSS3]{essAbsProd};
				
			\coordinate (c2) at (rel axis cs:1,1);
		\end{groupplot}
		\coordinate (c3) at ($(c1)!.5!(c2)$);
		\node[below] at (c3 |- current bounding box.south) {\pgfplotslegendfromname{legess}};
	\end{tikzpicture}
	\caption{Toxin. \textcolor{black}{ESS for the equilibration (left) and production (right) phases of the simulations. The observable of interest is the distance traversed by the toxin from its initial position.} Comparison of the two-stage \mbox{(M-)BCSS2}, (M-)ME(2), three-stage (M-)BCSS3, M-ME3, and Verlet integrators.}\label{IACFToxin}
\end{figure}

\textcolor{black}{Though acceptance rates contribute to the overall sampling efficiency of
 MHMC \cite{Akhmatskaya08}, the MHMC sampling performance cannot be determined by the acceptance rate on its own.}
   We shall see next how the choice of integrator impacts the sampling efficiency of GSHMC,
    \textcolor{black}{measured in terms of the number of uncorrelated samples, i.e.\  ESS, delivered by the different
    simulations.}
\textcolor{black}{Figure~\ref{IACFToxin} presents  ESS of the toxin drift to the preferred interfacial
location provided by GSHMC simulations using different integrators and time steps. In the left-hand graph,
  ESS is calculated for the equilibration phase. In the right-hand graph,  ESS is shown for the production
  phase of the simulation. Clearly, M-BCSS3 provides the best sampling, as measured by  ESS.
  For the largest time-step \(h=25\) fs Verlet leads to the ESS value slightly better than that of M-BCSS3.
  However, that largest value of \(h\) cannot be recommended on efficiency grounds, as 
  reducing the time-step  from \(h=25\) fs to \(h= 15\) fs increases the number of independent samples by a factor larger than three and thus makes the latter (h = 15 fs) a more appropriate choice for this application. }

\begin{figure}[!ht]
	\centering
	\begin{tikzpicture}
		\begin{groupplot}[group style={group size=2 by 1, horizontal sep=1.7cm, vertical sep=2cm}]
			\nextgroupplot[width=8.1cm, height=5.67cm,
				xmin=9.5, xmax=25.5, xlabel={$h$ (fs)}, xtick={10,15,20,25}, xticklabels={10,15,20,25},
				ymin=0, ymax=60, ytick={0,20,40,60}, yticklabels={0,20,40,60}, ylabel={$ESS$ (equilibration)},
				legend style={at={(0,0)},draw=black,fill=white,legend cell align=left, anchor=south},legend columns=4,legend to name=legess]
								
				\addplot[color=mygrey,mark=*,line width=2pt] table[x=x,y=V]{essAbsCoulomb};
				\addlegendentry{V};
								
				\addplot[color=myblue,dashed,mark=*,mark options={fill=white,solid},line width=1pt] table[x=x,y=ME]{essAbsCoulomb};
				\addlegendentry{ME};
							
				\addplot[color=myred,dashed,mark=*,mark options={fill=white,solid},line width=1pt] table[x=x,y=BCSS2]{essAbsCoulomb};
				\addlegendentry{BCSS2};
								
				\addplot[color=myred,solid,mark=*,line width=1pt] table[x=x,y=BCSS3]{essAbsCoulomb};
				\addlegendentry{BCSS3};
							
				\addplot[color=myblue,dashed,mark=*,mark options={fill=white,solid},line width=2pt] table[x=x,y=MME2]{essAbsCoulomb};
				\addlegendentry{M-ME2};
								
				\addplot[color=myblue,solid,mark=*,line width=2pt] table[x=x,y=MME3]{essAbsCoulomb};
				\addlegendentry{M-ME3};
								
				\addplot[color=myred,dashed,mark=*,mark options={fill=white,solid},line width=2pt] table[x=x,y=MBCSS2]{essAbsCoulomb};
				\addlegendentry{M-BCSS2};
								
				\addplot[color=myred,solid,mark=*,line width=2pt] table[x=x,y=MBCSS3]{essAbsCoulomb};
				\addlegendentry{M-BCSS3};
				
			\coordinate (c1) at (rel axis cs:0,1);

			\nextgroupplot[width=8.2cm, height=5.74cm,
				xmin=9.5, xmax=25.5, xlabel={$h$ (fs)}, xtick={10,15,20,25}, xticklabels={10,15,20,25},
				ymin=0, ymax=140, ytick={0,20,40,60,80,100,120,140}, yticklabels={0,20,40,60,80,100,120,140}, ylabel={$ESS$ (production)}]
								
				\addplot [color=mygrey, mark=*, line width=2pt] table[x=x,y=V]{essAbsProdCoulomb};
								
				\addplot [color=myblue,dashed,mark=*,mark options={fill=white,solid},line width=1pt] table[x=x,y=ME]{essAbsProdCoulomb};
							
				\addplot [color=myred,dashed,mark=*,mark options={fill=white,solid},line width=1pt] table[x=x,y=BCSS2]{essAbsProdCoulomb};
								
				\addplot [color=myred,solid,mark=*,line width=1pt] table[x=x,y=BCSS3]{essAbsProdCoulomb};
							
				\addplot [color=myblue,dashed,mark=*,mark options={fill=white,solid},line width=2pt] table[x=x,y=MME2]{essAbsProdCoulomb};
								
				\addplot [color=myblue,solid,mark=*,line width=2pt] table[x=x,y=MME3]{essAbsProdCoulomb};
								
				\addplot [color=myred,dashed,mark=*,mark options={fill=white,solid},line width=2pt] table[x=x,y=MBCSS2]{essAbsProdCoulomb};
								
				\addplot [color=myred,solid,mark=*,line width=2pt] table[x=x,y=MBCSS3]{essAbsProdCoulomb};
				
			\coordinate (c2) at (rel axis cs:1,1);
		\end{groupplot}
		\coordinate (c3) at ($(c1)!.5!(c2)$);
		\node[below] at (c3 |- current bounding box.south) {\pgfplotslegendfromname{legess}};
	\end{tikzpicture}
	\caption{\textcolor{black}{Toxin. ESS for the equilibration (left) and production (right) phases of the simulations. The observable of interest is the electrostatic/Coulombic interaction energy between the toxin and the bilayer. Comparison of the two-stage \mbox{(M-)BCSS2}, (M-)ME(2), three-stage (M-)BCSS3, M-ME3, and Verlet integrators.}}\label{CoulombToxin}
\end{figure}
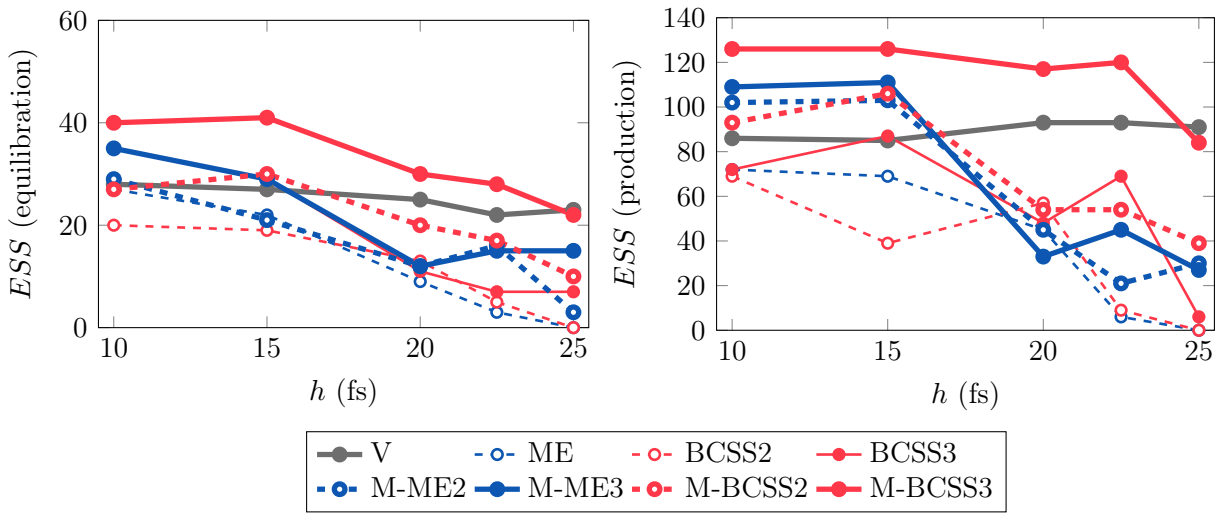

\textcolor{black}{The value of  ESS in a given simulation depends on the specific observable of interest. In
order  to check that the situation depicted in Figure~\ref{IACFToxin} for the toxin drift is not specific to
that observable, we have also studied ESS of the electrostatic/Coulombic interaction
 energy between the toxin and the bilayer. The results are given in Figure~\ref{CoulombToxin}, where,
 as in Figure~\ref{IACFToxin}, the left-hand graph corresponds to
   the equilibration phase and the right-hand graph to the
    production phase of the simulation. As for the toxin drift,
  M-BCSS3 provides the best sampling which is achieved at \(h=15\)~fs.}


\subsubsection{Multivariate Gaussian distribution}
In this section we test the performance of multi-stage integrators in MMHMC with the modified Hamiltonian \eqref{modHam4_an} used for sampling of multivariate Gaussian distributions.

The goal of this benchmark, proposed in \citep{HoffmanGelman14}, is to  sample from a $D$-dimensional Gaussian $\mathcal N(0,\Sigma)$, where the precision matrix $\Sigma^{-1}$ is generated from a Wishart distribution with $D$ degrees of freedom and the $D$-dimensional identity scale matrix.

We compare the performance of the standard Verlet integrator, and multi-stage schemes derived for problems with a quadratic potential, summarized in Table \ref{Tab:IntSummary}, for sampling from a multivariate Gaussian distribution of dimensions $D=100,1000,2000$.
We have identified a range of reasonable  values for the parameters $L$, $h$ and $\varphi$ and performed the comparisons for this range. For each MC iteration  the number of integration steps is drawn randomly from a uniform distribution  on $\{1,\dots,L \}$ and the step size is  uniformly distributed from $(0.8h, 1.2h)$. We report only results obtained with those choices of $\varphi$ and $L$ that provided the
 best performance of MMHMC regardless of the choice of  integrator. All the experiments here are carried out with the identity mass
matrix for MMHMC.
The number of production samples generated is $10000$ after a warm-up of $2000$ discarded samples.

\begin{figure}[!ht]
	\centering
	\begin{tikzpicture}
		\begin{groupplot}[group style={group size=3 by 1, horizontal sep=0.5cm, vertical sep=2cm}]
			\nextgroupplot[width=6.25cm, height=4.8438cm, title={$D = 100$},
				xmin=0.018333, xmax=0.071667, xlabel={$h$}, xtick={0.02,0.04,0.06}, xticklabels={0.02,0.04,0.06},
				ymin=40, ymax=100, ytick={25,50,75,100}, yticklabels={25,50,75,100}, ylabel={$AR$ (\%)}, scaled ticks=false,
				legend style={at={(1.44,-0.68)},draw=black,fill=white,legend cell align=left, anchor=south}, legend columns=4, legend to name=legars]
		
				\addplot[color=mygrey,mark=*,line width=2pt] table[x=x,y=V]{ar_100.dat};
				\addlegendentry{V};
		
				\addplot[color=myblue,dashed,mark=*,mark options={fill=white,solid},line width=1pt] table[x=x,y=ME]{ar_100.dat};
				\addlegendentry{ME};
		
				\addplot[color=myred,dashed,mark=*,mark options={fill=white,solid},line width=1pt] table[x=x,y=BCSS2]{ar_100.dat};
				\addlegendentry{BCSS2};
		
				\addplot[color=myred,solid,mark=*,line width=1pt] table[x=x,y=BCSS3]{ar_100.dat};
				\addlegendentry{BCSS3};
		
				\addplot[color=myblue,dashed,mark=*,mark options={fill=white,solid},line width=2pt] table[x=x,y=MME2]{ar_100.dat};
				\addlegendentry{M-ME2};
		
				\addplot[color=myblue,solid,mark=*,line width=2pt] table[x=x,y=MME3]{ar_100.dat};
				\addlegendentry{M-ME3};
		
				\addplot[color=myred,dashed,mark=*,mark options={fill=white,solid},line width=2pt] table[x=x,y=MBCSS2]{ar_100.dat};
				\addlegendentry{M-BCSS2};
		
				\addplot[color=myred,solid,mark=*,line width=2pt] table[x=x,y=MBCSS3]{ar_100.dat};
				\addlegendentry{M-BCSS3};
				
			\coordinate (c1) at (rel axis cs:0,1);
		
			\nextgroupplot[width=6.25cm, height=4.8438cm, title={$D = 1000$},
				xmin=0.0056667, xmax=0.016333, xlabel={$h$}, xtick={0.006,0.010,0.014}, xticklabels={0.006,0.010,0.014},
				ymin=40, ymax=100, ytick={25,50,75,100}, yticklabels={}, scaled ticks=false]
	
				\addplot[color=mygrey,mark=*,line width=2pt] table[x=x,y=V]{ar_1000.dat};
		
				\addplot[color=myblue,dashed,mark=*,mark options={fill=white,solid},line width=1pt] table[x=x,y=ME]{ar_1000.dat};
		
				\addplot[color=myred,dashed,mark=*,mark options={fill=white,solid},line width=1pt] table[x=x,y=BCSS2]{ar_1000.dat};
		
				\addplot[color=myred,solid,mark=*,line width=1pt] table[x=x,y=BCSS3]{ar_1000.dat};
		
				\addplot[color=myblue,dashed,mark=*,mark options={fill=white,solid},line width=2pt] table[x=x,y=MME2]{ar_1000.dat};
		
				\addplot[color=myblue,solid,mark=*,line width=2pt] table[x=x,y=MME3]{ar_1000.dat};
		
				\addplot[color=myred,dashed,mark=*,mark options={fill=white,solid},line width=2pt] table[x=x,y=MBCSS2]{ar_1000.dat};
		
				\addplot[color=myred,solid,mark=*,line width=2pt] table[x=x,y=MBCSS3]{ar_1000.dat};
		
			\nextgroupplot[width=6.25cm, height=4.8438cm, title={$D = 2000$},
				xmin=0.0038333, xmax=0.0091667, xlabel={$h$}, xtick={0.004,0.006,0.008}, xticklabels={0.004,0.006,0.008},
				ymin=40, ymax=100, ylabel={}, ytick={25,50,75,100}, yticklabels={}, scaled ticks=false]
	
				\addplot[restrict x to domain=0.004:0.009][color=mygrey,mark=*,line width=2pt] table[x=x,y=V]{ar_2000.dat};
	
				\addplot[restrict x to domain=0.004:0.009][color=myblue,dashed,mark=*,mark options={fill=white,solid},line width=1pt] table[x=x,y=ME]{ar_2000.dat};
	
				\addplot[restrict x to domain=0.004:0.009][color=myred,dashed,mark=*,mark options={fill=white,solid},line width=1pt] table[x=x,y=BCSS2]{ar_2000.dat};
	
				\addplot[restrict x to domain=0.004:0.009][color=myred,solid,mark=*,line width=1pt] table[x=x,y=BCSS3]{ar_2000.dat};
	
				\addplot[restrict x to domain=0.004:0.009][color=myblue,dashed,mark=*,mark options={fill=white,solid},line width=2pt] table[x=x,y=MME2]{ar_2000.dat};
	
				\addplot[restrict x to domain=0.004:0.009][color=myblue,solid,mark=*,line width=2pt] table[x=x,y=MME3]{ar_2000.dat};
	
				\addplot[restrict x to domain=0.004:0.009][color=myred,dashed,mark=*,mark options={fill=white,solid},line width=2pt] table[x=x,y=MBCSS2]{ar_2000.dat};
	
				\addplot[restrict x to domain=0.004:0.009][color=myred,solid,mark=*,line width=2pt] table[x=x,y=MBCSS3]{ar_2000.dat};
				
			\coordinate (c2) at (rel axis cs:1,1);
		\end{groupplot}
		\coordinate (c3) at ($(c1)!.5!(c2)$);
		\node[below] at (c3 |- current bounding box.south) {\pgfplotslegendfromname{legars}};
	\end{tikzpicture}
	\caption{Acceptance rates as functions of the step size $h$ for sampling from a $D$-dimensional Gaussian distribution. Comparison of the two-stage \mbox{(M-)BCSS2}, (M-)ME(2), three-stage (M-)BCSS3, M-ME3, and Verlet integrators.}
	\label{Fig:AR}
\end{figure}
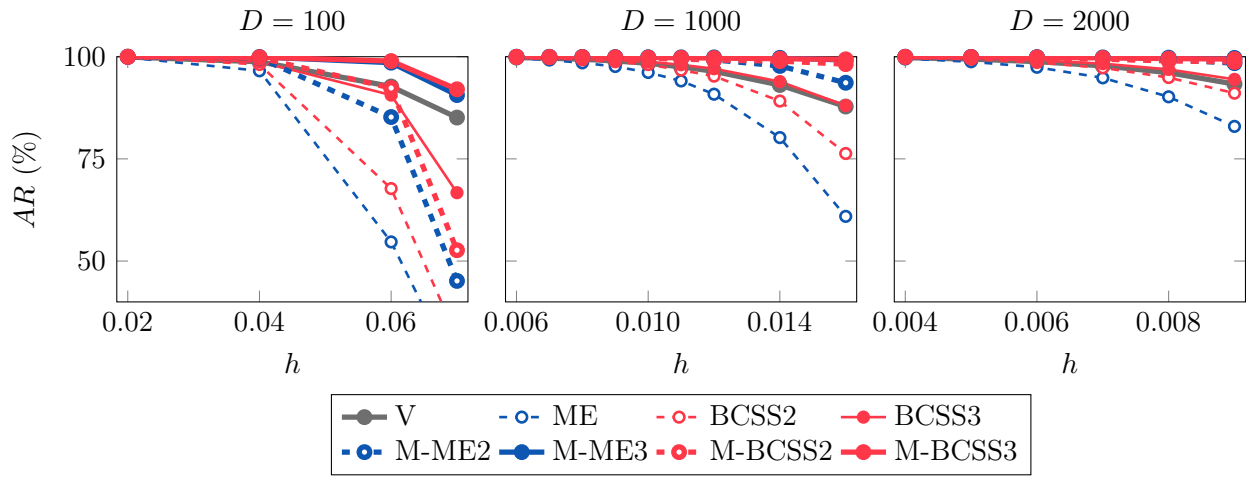

Figure \ref{Fig:AR} presents the resulting acceptance rates as functions of the step size $h$. MHMC-specific
integrators always lead to higher AR than their counterparts derived for the HMC method. We note that for the
smallest dimension ($D=100$) the Verlet integrator 
\textcolor{black}{provides higher acceptance rates} than all two-stage integrators, due to its
larger stability limit. However, the new three-stage methods outperform Verlet even for this dimension. For
bigger dimensions, which require smaller step sizes, the better conservation of the modified Hamiltonian by
all multi-stage integrators (see Figure \ref{Rho_func}) implies higher acceptance rates. In this case, all the
newly derived multi-stage integrators show improvement over Verlet, with M-BCSS3 
\textcolor{black}{being the} best.

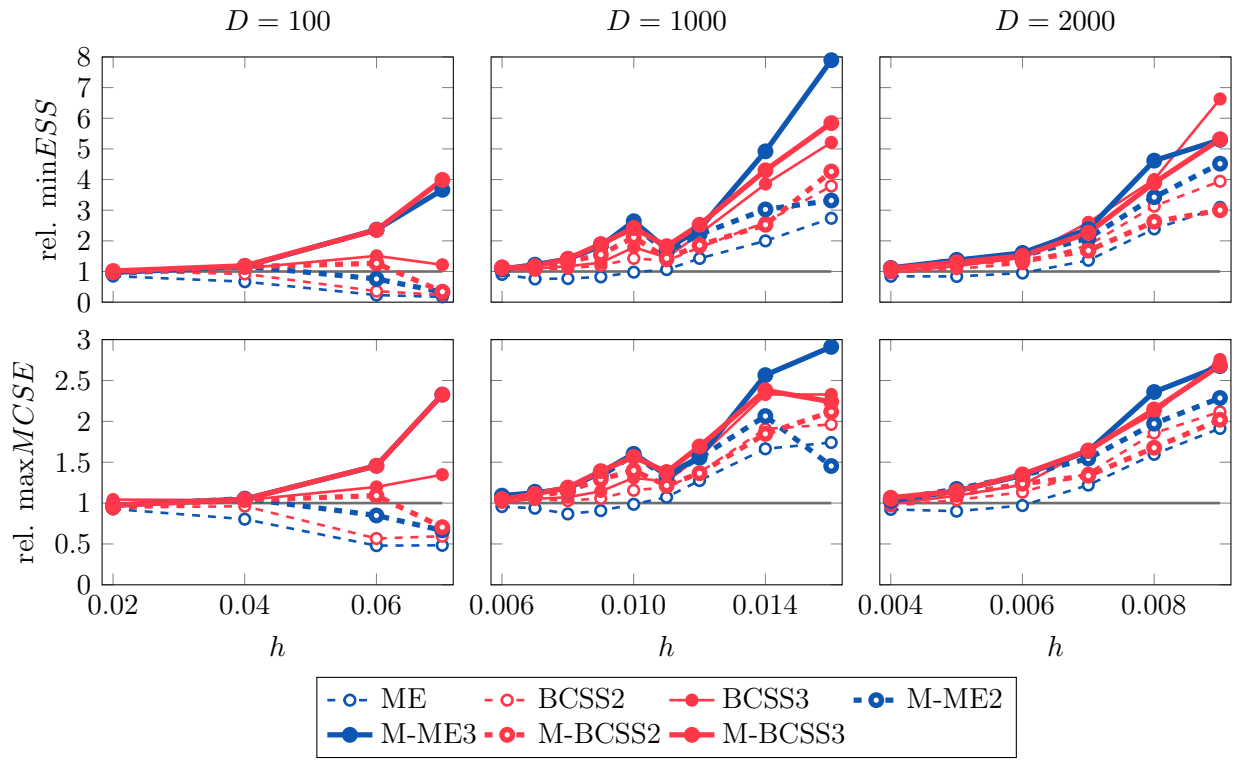
\begin{figure}[!ht]
	\begin{tikzpicture}
		\begin{groupplot}[group style={group size=3 by 2, horizontal sep=0.5cm, vertical sep=0.5cm}]
			\nextgroupplot[width=6.25cm, height=4.8438cm, title={$D = 100$},
				xmin=0.018333, xmax=0.071667, xlabel={}, xtick={0.02,0.04,0.06}, xticklabels={},
				ymin=0, ymax=8, ytick={0,1,2,3,4,5,6,7,8}, yticklabels={0,1,2,3,4,5,6,7,8}, ylabel={rel. min$ESS$}, scaled ticks=false,
				legend style={at={(0,0)},draw=black,fill=white,legend cell align=left, anchor=south}, legend columns=4, legend to name=legsampling]
	
				\addplot+[color=mygrey,mark=none,line width=1pt,forget plot] table[x=x,y expr={\thisrow{V}/\thisrow{V}}]{ess_100.dat};
				
				\addplot[color=myblue,dashed,mark=*,mark options={fill=white,solid},line width=1pt] table[x=x,y expr={\thisrow{ME}/\thisrow{V}}]{ess_100.dat};
				\addlegendentry{ME};
	
				\addplot[color=myred,dashed,mark=*,mark options={fill=white,solid},line width=1pt] table[x=x,y expr={\thisrow{BCSS2}/\thisrow{V}}]{ess_100.dat};
				\addlegendentry{BCSS2};
	
				\addplot[color=myred,solid,mark=*,line width=1pt] table[x=x,y expr={\thisrow{BCSS3}/\thisrow{V}}]{ess_100.dat};
				\addlegendentry{BCSS3};
	
				\addplot[color=myblue,dashed,mark=*,mark options={fill=white,solid},line width=2pt] table[x=x,y expr={\thisrow{MME2}/\thisrow{V}}]{ess_100.dat};
				\addlegendentry{M-ME2};
	
				\addplot[color=myblue,solid,mark=*,line width=2pt] table[x=x,y expr={\thisrow{MME3}/\thisrow{V}}]{ess_100.dat};
				\addlegendentry{M-ME3};
	
				\addplot[color=myred,dashed,mark=*,mark options={fill=white,solid},line width=2pt] table[x=x,y expr={\thisrow{MBCSS2}/\thisrow{V}}]{ess_100.dat};
				\addlegendentry{M-BCSS2};
	
				\addplot[color=myred,solid,mark=*,line width=2pt] table[x=x,y expr={\thisrow{MBCSS3}/\thisrow{V}}]{ess_100.dat};
				\addlegendentry{M-BCSS3};

			\nextgroupplot[width=6.25cm, height=4.8438cm, title={$D = 1000$},
				xmin=0.0056667, xmax=0.016333, xlabel={}, xtick={0.006,0.010,0.014}, xticklabels={},
				ymin=0, ymax=8, ytick={0,1,2,3,4,5,6,7,8}, yticklabels={}, scaled ticks=false]
	
				\addplot+[color=mygrey,mark=none,line width=1pt,forget plot] table[x=x,y expr={\thisrow{V}/\thisrow{V}}]{ess_1000.dat};
	
				\addplot[color=myblue,dashed,mark=*,mark options={fill=white,solid},line width=1pt] table[x=x,y expr={\thisrow{ME}/\thisrow{V}}]{ess_1000.dat};
	
				\addplot[color=myred,dashed,mark=*,mark options={fill=white,solid},line width=1pt] table[x=x,y expr={\thisrow{BCSS2}/\thisrow{V}}]{ess_1000.dat};
	
				\addplot[color=myred,solid,mark=*,line width=1pt] table[x=x,y expr={\thisrow{BCSS3}/\thisrow{V}}]{ess_1000.dat};
	
				\addplot[color=myblue,dashed,mark=*,mark options={fill=white,solid},line width=2pt] table[x=x,y expr={\thisrow{MME2}/\thisrow{V}}]{ess_1000.dat};
	
				\addplot[color=myblue,solid,mark=*,line width=2pt] table[x=x,y expr={\thisrow{MME3}/\thisrow{V}}]{ess_1000.dat};
	
				\addplot[color=myred,dashed,mark=*,mark options={fill=white,solid},line width=2pt] table[x=x,y expr={\thisrow{MBCSS2}/\thisrow{V}}]{ess_1000.dat};
	
				\addplot[color=myred,solid,mark=*,line width=2pt] table[x=x,y expr={\thisrow{MBCSS3}/\thisrow{V}}]{ess_1000.dat};
		
			\nextgroupplot[width=6.25cm, height=4.8438cm, title={$D = 2000$},
				xmin=0.0038333, xmax=0.0091667, xlabel={}, xtick={0.004,0.006,0.008}, xticklabels={},
				ymin=0, ymax=8, ytick={0,1,2,3,4,5,6,7,8}, yticklabels={}, scaled ticks=false]
	
				\addplot+[restrict x to domain=0.004:0.009][color=mygrey,mark=none,line width=1pt,forget plot] table[x=x,y expr={\thisrow{V}/\thisrow{V}}]{ess_2000.dat};
			
				\addplot[restrict x to domain=0.004:0.009][color=myblue,dashed,mark=*,mark options={fill=white,solid},line width=1pt] table[x=x,y expr={\thisrow{ME}/\thisrow{V}}]{ess_2000.dat};
			
				\addplot[restrict x to domain=0.004:0.009][color=myred,dashed,mark=*,mark options={fill=white,solid},line width=1pt] table[x=x,y expr={\thisrow{BCSS2}/\thisrow{V}}]{ess_2000.dat};
			
				\addplot[restrict x to domain=0.004:0.009][color=myred,solid,mark=*,line width=1pt] table[x=x,y expr={\thisrow{BCSS3}/\thisrow{V}}]{ess_2000.dat};
			
				\addplot[restrict x to domain=0.004:0.009][color=myblue,dashed,mark=*,mark options={fill=white,solid},line width=2pt] table[x=x,y expr={\thisrow{MME2}/\thisrow{V}}]{ess_2000.dat};
			
				\addplot[restrict x to domain=0.004:0.009][color=myblue,solid,mark=*,line width=2pt] table[x=x,y expr={\thisrow{MME3}/\thisrow{V}}]{ess_2000.dat};
			
				\addplot[restrict x to domain=0.004:0.009][color=myred,dashed,mark=*,mark options={fill=white,solid},line width=2pt] table[x=x,y expr={\thisrow{MBCSS2}/\thisrow{V}}]{ess_2000.dat};
			
				\addplot[restrict x to domain=0.004:0.009][color=myred,solid,mark=*,line width=2pt] table[x=x,y expr={\thisrow{MBCSS3}/\thisrow{V}}]{ess_2000.dat};
		
			\nextgroupplot[width=6.25cm, height=4.8438cm,
				xmin=0.018333, xmax=0.071667, xlabel={$h$}, xtick={0.02,0.04,0.06}, xticklabels={0.02,0.04,0.06},
				ymin=0, ymax=3, ytick={0,0.5,1,1.5,2,2.5,3}, yticklabels={0,0.5,1,1.5,2,2.5,3}, ylabel={rel. max$MCSE$}, scaled ticks=false]
	
				\addplot+[color=mygrey,mark=none,line width=1pt,forget plot] table[x=x,y expr={\thisrow{V}/\thisrow{V}}]{mcse_100.dat};
		
				\addplot[color=myblue,dashed,mark=*,mark options={fill=white,solid},line width=1pt] table[x=x,y expr={\thisrow{V}/\thisrow{ME}}]{mcse_100.dat};
		
				\addplot[color=myred,dashed,mark=*,mark options={fill=white,solid},line width=1pt] table[x=x,y expr={\thisrow{V}/\thisrow{BCSS2}}]{mcse_100.dat};
		
				\addplot[color=myred,solid,mark=*,line width=1pt] table[x=x,y expr={\thisrow{V}/\thisrow{BCSS3}}]{mcse_100.dat};
		
				\addplot[color=myblue,dashed,mark=*,mark options={fill=white,solid},line width=2pt] table[x=x,y expr={\thisrow{V}/\thisrow{MME2}}]{mcse_100.dat};
		
				\addplot[color=myblue,solid,mark=*,line width=2pt] table[x=x,y expr={\thisrow{V}/\thisrow{MME3}}]{mcse_100.dat};
		
				\addplot[color=myred,dashed,mark=*,mark options={fill=white,solid},line width=2pt] table[x=x,y expr={\thisrow{V}/\thisrow{MBCSS2}}]{mcse_100.dat};
		
				\addplot[color=myred,solid,mark=*,line width=2pt] table[x=x,y expr={\thisrow{V}/\thisrow{MBCSS3}}]{mcse_100.dat};
				
			\coordinate (c1) at (rel axis cs:0,1);
		
			\nextgroupplot[width=6.25cm, height=4.8438cm,
				xmin=0.0056667, xmax=0.016333, xlabel={$h$}, xtick={0.006,0.010,0.014}, xticklabels={0.006,0.010,0.014},
				ymin=0, ymax=3, ytick={0,0.5,1,1.5,2,2.5,3}, yticklabels={}, scaled ticks=false]
	
				\addplot+[color=mygrey,mark=none,line width=1pt,forget plot] table[x=x,y expr={\thisrow{V}/\thisrow{V}}]{mcse_1000.dat};
		
				\addplot[color=myblue,dashed,mark=*,mark options={fill=white,solid},line width=1pt] table[x=x,y expr={\thisrow{V}/\thisrow{ME}}]{mcse_1000.dat};
		
				\addplot[color=myred,dashed,mark=*,mark options={fill=white,solid},line width=1pt] table[x=x,y expr={\thisrow{V}/\thisrow{BCSS2}}]{mcse_1000.dat};
		
				\addplot[color=myred,solid,mark=*,line width=1pt] table[x=x,y expr={\thisrow{V}/\thisrow{BCSS3}}]{mcse_1000.dat};
		
				\addplot[color=myblue,dashed,mark=*,mark options={fill=white,solid},line width=2pt] table[x=x,y expr={\thisrow{V}/\thisrow{MME2}}]{mcse_1000.dat};
		
				\addplot[color=myblue,solid,mark=*,line width=2pt] table[x=x,y expr={\thisrow{V}/\thisrow{MME3}}]{mcse_1000.dat};
		
				\addplot[color=myred,dashed,mark=*,mark options={fill=white,solid},line width=2pt] table[x=x,y expr={\thisrow{V}/\thisrow{MBCSS2}}]{mcse_1000.dat};
		
				\addplot[color=myred,solid,mark=*,line width=2pt] table[x=x,y expr={\thisrow{V}/\thisrow{MBCSS3}}]{mcse_1000.dat};
		
			\nextgroupplot[width=6.25cm, height=4.8438cm,
				xmin=0.0038333, xmax=0.0091667, xlabel={$h$}, xtick={0.004,0.006,0.008}, xticklabels={0.004,0.006,0.008},
				ymin=0, ymax=3, ytick={0,0.5,1,1.5,2,2.5,3}, yticklabels={}, scaled ticks=false]
	
				\addplot+[restrict x to domain=0.004:0.009][color=mygrey,mark=none,line width=1pt,forget plot] table[x=x,y expr={\thisrow{V}/\thisrow{V}}]{mcse_2000.dat};
		
				\addplot[restrict x to domain=0.004:0.009][color=myblue,dashed,mark=*,mark options={fill=white,solid},line width=1pt] table[x=x,y expr={\thisrow{V}/\thisrow{ME}}]{mcse_2000.dat};
		
				\addplot[restrict x to domain=0.004:0.009][color=myred,dashed,mark=*,mark options={fill=white,solid},line width=1pt] table[x=x,y expr={\thisrow{V}/\thisrow{BCSS2}}]{mcse_2000.dat};
		
				\addplot[restrict x to domain=0.004:0.009][color=myred,solid,mark=*,line width=1pt] table[x=x,y expr={\thisrow{V}/\thisrow{BCSS3}}]{mcse_2000.dat};
		
				\addplot[restrict x to domain=0.004:0.009][color=myblue,dashed,mark=*,mark options={fill=white,solid},line width=2pt] table[x=x,y expr={\thisrow{V}/\thisrow{MME2}}]{mcse_2000.dat};
		
				\addplot[restrict x to domain=0.004:0.009][color=myblue,solid,mark=*,line width=2pt] table[x=x,y expr={\thisrow{V}/\thisrow{MME3}}]{mcse_2000.dat};
		
				\addplot[restrict x to domain=0.004:0.009][color=myred,dashed,mark=*,mark options={fill=white,solid},line width=2pt] table[x=x,y expr={\thisrow{V}/\thisrow{MBCSS2}}]{mcse_2000.dat};
		
				\addplot[restrict x to domain=0.004:0.009][color=myred,solid,mark=*,line width=2pt] table[x=x,y expr={\thisrow{V}/\thisrow{MBCSS3}}]{mcse_2000.dat};
			
			\coordinate (c2) at (rel axis cs:1,1);
		\end{groupplot}
		\coordinate (c3) at ($(c1)!.5!(c2)$);
		\node[below] at (c3 |- current bounding box.south) {\pgfplotslegendfromname{legsampling}};
	\end{tikzpicture}
	\caption{Relative sampling performance with respect to the Verlet integrator, in terms of minimum ESS (top) and maximum MCSE (bottom) over variates, as functions of the step size $h$ for sampling from a $D$-dimensional Gaussian distribution. Comparison of the two-stage \mbox{(M-)BCSS2}, (M-)ME(2), three-stage (M-)BCSS3, M-ME3, and Verlet integrators.}
	\label{Fig:ESS}
\end{figure}

\begin{figure}[!ht]
	\centering
	\begin{tikzpicture}
		\begin{groupplot}[group style={group size=3 by 1, horizontal sep=0.5cm, vertical sep=2cm}]
			\nextgroupplot[width=6.25cm, height=4.8438cm, title={$D = 100$},
				xmin=0.018333, xmax=0.071667, xlabel={$h$}, xtick={0.02,0.04,0.06}, xticklabels={0.02,0.04,0.06},
				ymin=0, ymax=30, ytick={0,5,10,15,20,25,30}, yticklabels={0,5,10,15,20,25,30}, ylabel={total dist.\ from the mean}, scaled ticks=false,
				legend style={at={(1.48,-0.7)},draw=black,fill=white,legend cell align=left, anchor=south}, legend columns=4, legend to name=legdistance]
	
				\addplot[color=mygrey,mark=*,line width=2pt] table[x=x,y=V]{d_100.dat};
				\addlegendentry{V};
	
				\addplot[color=myblue,dashed,mark=*,mark options={fill=white,solid},line width=1pt] table[x=x,y=ME]{d_100.dat};
				\addlegendentry{ME};
	
				\addplot[color=myred,dashed,mark=*,mark options={fill=white,solid},line width=1pt] table[x=x,y=BCSS2]{d_100.dat};
				\addlegendentry{BCSS2};
	
				\addplot[color=myred,solid,mark=*,line width=1pt] table[x=x,y=BCSS3]{d_100.dat};
				\addlegendentry{BCSS3};
	
				\addplot[color=myblue,dashed,mark=*,mark options={fill=white,solid},line width=2pt] table[x=x,y=MME2]{d_100.dat};
				\addlegendentry{M-ME2};
	
				\addplot[color=myblue,solid,mark=*,line width=2pt] table[x=x,y=MME3]{d_100.dat};
				\addlegendentry{M-ME3};
	
				\addplot[color=myred,dashed,mark=*,mark options={fill=white,solid},line width=2pt] table[x=x,y=MBCSS2]{d_100.dat};
				\addlegendentry{M-BCSS2};
	
				\addplot[color=myred,solid,mark=*,line width=2pt] table[x=x,y=MBCSS3]{d_100.dat};
				\addlegendentry{M-BCSS3};
				
			\coordinate (c1) at (rel axis cs:0,1);
		
			\nextgroupplot[width=6.25cm, height=4.8438cm, title={$D = 1000$},
				xmin=0.0056667, xmax=0.016333, xlabel={$h$}, xtick={0.006,0.010,0.014}, xticklabels={0.006,0.010,0.014},
				ymin=0, ymax=30, ytick={0,5,10,15,20,25,30}, yticklabels={}, scaled ticks=false]
	
				\addplot[color=mygrey,mark=*,line width=2pt] table[x=x,y=V]{d_1000.dat};
	
				\addplot[color=myblue,dashed,mark=*,mark options={fill=white,solid},line width=1pt] table[x=x,y=ME]{d_1000.dat};
	
				\addplot[color=myred,dashed,mark=*,mark options={fill=white,solid},line width=1pt] table[x=x,y=BCSS2]{d_1000.dat};
	
				\addplot[color=myred,solid,mark=*,line width=1pt] table[x=x,y=BCSS3]{d_1000.dat};
	
				\addplot[color=myblue,dashed,mark=*,mark options={fill=white,solid},line width=2pt] table[x=x,y=MME2]{d_1000.dat};
	
				\addplot[color=myblue,solid,mark=*,line width=2pt] table[x=x,y=MME3]{d_1000.dat};
	
				\addplot[color=myred,dashed,mark=*,mark options={fill=white,solid},line width=2pt] table[x=x,y=MBCSS2]{d_1000.dat};
	
				\addplot[color=myred,solid,mark=*,line width=2pt] table[x=x,y=MBCSS3]{d_1000.dat};
		
			\nextgroupplot[width=6.25cm, height=4.8438cm, title={$D = 2000$},
				xmin=0.0038333, xmax=0.0091667, xlabel={$h$}, xtick={0.004,0.006,0.008}, xticklabels={0.004,0.006,0.008},
				ymin=0, ymax=30, ytick={0,5,10,15,20,25,30}, yticklabels={}, scaled ticks=false]
	
				\addplot[restrict x to domain=0.004:0.009][color=mygrey,mark=*,line width=2pt] table[x=x,y=V]{d_2000.dat};
	
				\addplot[restrict x to domain=0.004:0.009][color=myblue,dashed,mark=*,mark options={fill=white,solid},line width=1pt] table[x=x,y=ME]{d_2000.dat};
	
				\addplot[restrict x to domain=0.004:0.009][color=myred,dashed,mark=*,mark options={fill=white,solid},line width=1pt] table[x=x,y=BCSS2]{d_2000.dat};
	
				\addplot[restrict x to domain=0.004:0.009][color=myred,solid,mark=*,line width=1pt] table[x=x,y=BCSS3]{d_2000.dat};
	
				\addplot[restrict x to domain=0.004:0.009][color=myblue,dashed,mark=*,mark options={fill=white,solid},line width=2pt] table[x=x,y=MME2]{d_2000.dat};
	
				\addplot[restrict x to domain=0.004:0.009][color=myblue,solid,mark=*,line width=2pt] table[x=x,y=MME3]{d_2000.dat};
	
				\addplot[restrict x to domain=0.004:0.009][color=myred,dashed,mark=*,mark options={fill=white,solid},line width=2pt] table[x=x,y=MBCSS2]{d_2000.dat};
	
				\addplot[restrict x to domain=0.004:0.009][color=myred,solid,mark=*,line width=2pt] table[x=x,y=MBCSS3]{d_2000.dat};
				
			\coordinate (c2) at (rel axis cs:1,1);
		\end{groupplot}
		\coordinate (c3) at ($(c1)!.5!(c2)$);
		\node[below] at (c3 |- current bounding box.south) {\pgfplotslegendfromname{legdistance}};
	\end{tikzpicture}
	\caption{Total distance from the mean 	as function of the step size $h$ for sampling from a $D$-dimensional Gaussian distribution. Comparison of the two-stage \mbox{(M-)BCSS2}, (M-)ME(2), three-stage (M-)BCSS3, M-ME3, and Verlet integrators.}
	\label{Fig:Dist}
\end{figure}
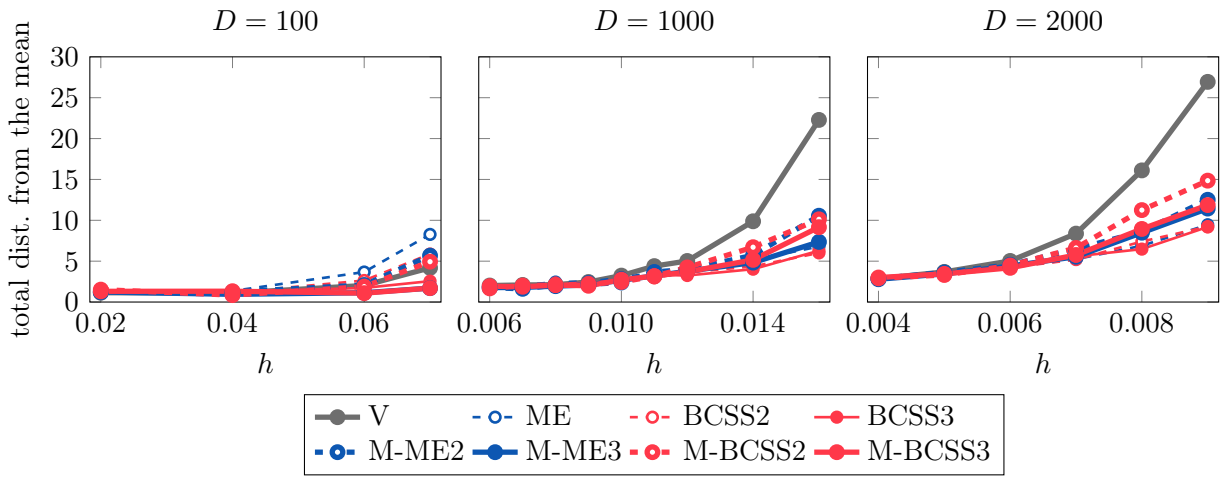

The relative sampling performance with respect to the Verlet integrator, in terms of minimum ESS (top)
 and maximum MCSE (bottom) over variates for the mean estimate (i.e.\ $\min_{d\in D}ESS(\hat{x}_d)$ and $\max_{d\in D}MCSE(\hat{x}_d)$),
 is presented in Figure \ref{Fig:ESS}. Values below 1 correspond to cases
  of sampling efficiency lower than Verlet's  and  values above 1 correspond to
   integrators that outperform  Verlet.
As in the case of the acceptance rates, for the smallest dimension, the Verlet integrator demonstrates a
better performance than all two-stage methods. We note that for the smallest step sizes there is no difference
among integrators. For bigger step sizes, the novel three-stage integrators outperform all other schemes and
improve sampling efficiency over  Verlet up to 8 and 3 times for ESS and MCSE, respectively. The improvement
increases with dimension; therefore we believe that for high dimensional problems the new multi-stage
integrators provide  crucial ingredients of  efficient samplers.
In addition, Figure \ref{Fig:Dist} presents a comparison in terms of total distance from the mean of the
target distribution;  lower values correspond to better accuracy.
For high dimensions, the MHMC-specific integrators demonstrate higher accuracy than the commonly used Verlet.

In summary, the tests performed on multivariate Gaussian distribution demonstrated the superiority of the integrators specifically designed for the MHMC methods  over both the standard Verlet and 
splitting schemes for the HMC method. The three-stage modified integrators clearly outperform all tested integrators for this problem.

\section{Conclusions}\label{Sec:Conclusions}

We have introduced  new multi-stage integrators for enhanced sampling with modified Hamiltonian Monte Carlo
(MHMC) methods. The proposed two- and three-stage integration methods provide, \textcolor{black}{for reasonable choices of time steps,} better conservation of modified
Hamiltonians than the Verlet integrator commonly used in MHMC. Each of the methods derived is characterized by
its coefficients, which were obtained from the minimization of the (expected) error in modified Hamiltonians
introduced by numerical integration. Accordingly, we proposed computationally efficient expressions for
modified Hamiltonians of order 4 and 6 for the multi-stage splitting integrating schemes. The new methods were
tested and compared with Verlet and also with 
splitting integrators previously suggested for
sampling with HMC. The comparisons use benchmarks in molecular simulation and computational statistics
problems, sampled with the  GSHMC \cite{Akhmatskaya08} and MMHMC \cite{Radivojevic:2017} methods,
respectively. Both method GSHMC and MMHMC belong to the MHMC class and had previously been shown to provide
good performance when sampling in applications envisaged here. The tests of the new integration schemes reveal
that the novel three-stage integrators lead to an outstanding improvement over the Verlet integrator for
problems in which the potential function is (approximately) quadratic. The improvement, which for the tested
systems is of up to 8 times, comes both in terms of acceptance rate and sampling efficiency over a range of
simulation parameters. For such application problems, all new integrators specifically derived for MHMC
methods outperform their counterparts proposed previously for HMC.

\section*{Acknowledgements}
The authors would like to thank the financial support from MTM2013-46553-C3-1-P
funded by the Spanish Ministry of Economy and Competitiveness (MINECO). They have also
received funding from the Project of MINECO with reference MTM2016-76329-R\break (AEI/FEDER, EU).
This work has been possible thanks to the support of the computing infrastructure of the i2BASQUE academic
network, and the technical and human support provided by IZO-SGI SGIker of UPV/EHU and European funding
(ERDF and ESF). E.A. and T.R. thank support from Basque Government - ELKARTEK Programme, grant KK-2016/0002.
 M.F.P. would like to thank MINECO for funding through the fellowship BES-2014-068640.
 J.M.S.S. has been additionally supported by project MTM2016-77660- P(AEI/FEDER, UE) funded by MINECO  and by project VA024P17, FEDER EU Junta de Castilla y Le\'on.
 This research is also supported by the Basque Government through the BERC 2014-2017 program and by MINECO:
 BCAM Severo Ochoa accreditation SEV-2013-0323.

\bibliography{MINT.bib}

\end{document}